%% file: manuscript.tex
\documentclass[twocolumn]{aastex61}
\pdfoutput=1 
\usepackage[utf8]{inputenc}
\usepackage{hyperref}
\usepackage{natbib}
\usepackage{graphicx}
\usepackage{xspace}
\usepackage[caption=false]{subfig}

\newcommand{\rsun}{$R_{\odot}$\xspace}
\newcommand{\mearth}{$M_\earth$\xspace}
\newcommand{\rearth}{$R_\earth$\xspace}

\newcommand{\rstar}{\ensuremath{R_{\star}}\xspace}

\newcommand{\feh}{\ensuremath{[\mbox{Fe}/\mbox{H}]}\xspace}
\newcommand{\teff}{\ensuremath{T_{\mathrm{eff}}}\xspace}
\newcommand{\logg}{\ensuremath{\log g}\xspace}
\newcommand{\vsini}{\ensuremath{v \sin i}\xspace}

\newcommand{\rp}{\ensuremath{R_p}\xspace}
\newcommand{\teq}{$T_{\mathrm{eq}}$\xspace}

\newcommand{\kepler}{\textit{Kepler}\xspace}
\newcommand{\ktwo}{\textit{K2}\xspace}
\newcommand{\spitzer}{\textit{Spitzer}\xspace}

\newcommand{\cheops}{\textit{CHEOPS}\xspace}

\newcommand{\medrad}{\ensuremath{2.2}\xspace}
\newcommand{\medper}{\ensuremath{6.9}\xspace}
\newcommand{\medteq}{\ensuremath{890}\xspace}
\newcommand{\medjmag}{\ensuremath{11.2}\xspace}

\newcommand{\numtotal}{72\xspace}
\newcommand{\numvalid}{44\xspace}
\newcommand{\numnew}{24\xspace}
\newcommand{\numcand}{27\xspace}

\newcommand{\nummulti}{18\xspace}
\newcommand{\nusp}{four\xspace}
\newcommand{\nsmall}{16\xspace}
\newcommand{\natmospheric}{two\xspace}

\shorttitle{{\it K2} Campaign 10}
\shortauthors{Livingston et al.}

\begin{document}

\title{\numvalid validated planets from \ktwo Campaign 10}

\author{John H. Livingston}
\affiliation{Department of Astronomy, University of Tokyo, 7-3-1 Hongo, Bunkyo-ku, Tokyo 113-0033, Japan}
\affiliation{JSPS Fellow}
\affiliation{\href{mailto:livingston@astron.s.u-tokyo.edu}{{\tt livingston@astron.s.u-tokyo.edu}}}

\author{Michael Endl}
\affiliation{Department of Astronomy and McDonald Observatory, University of Texas at Austin, 2515 Speedway,~Stop~C1400,~Austin,~TX~78712,~USA}

\author{Fei Dai}
\affiliation{Department of Physics and Kavli Institute for Astrophysics and Space Research, Massachusetts Institute of Technology, Cambridge, MA, 02139, USA}
\affiliation{Department of Astrophysical Sciences, Princeton University, 4 Ivy Lane, Princeton, NJ 08544, USA}

\author{William D. Cochran}
\affiliation{Department of Astronomy and McDonald Observatory, University of Texas at Austin, 2515 Speedway,~Stop~C1400,~Austin,~TX~78712,~USA}

\author{Oscar Barragan}
\affiliation{Dipartimento di Fisica, Universit\`a di Torino, via P. Giuria 1, 10125 Torino, Italy}

\author{Davide Gandolfi}
\affiliation{Dipartimento di Fisica, Universit\`a di Torino, via P. Giuria 1, 10125 Torino, Italy}

\author{Teruyuki Hirano}
\affiliation{Department of Earth and Planetary Sciences, Tokyo Institute of Technology, 2-12-1 Ookayama, Meguro-ku, Tokyo 152-8551, Japan}

\author{Sascha Grziwa}
\affiliation{Rheinisches Institut f\"ur Umweltforschung an der Universit\"at zu K\"oln, Aachener Strasse 209, 50931 K\"oln, Germany}

\author{Alexis M. S. Smith}
\affiliation{Institute of Planetary Research, German Aerospace Center, Rutherfordstrasse 2, 12489 Berlin, Germany}


\author{Simon Albrecht}
\affiliation{Stellar Astrophysics Centre, Department of Physics and Astronomy, Aarhus University, Ny Munkegade 120, DK-8000 Aarhus C, Denmark}

\author{Juan Cabrera}
\affiliation{Institute of Planetary Research, German Aerospace Center, Rutherfordstrasse 2, 12489 Berlin, Germany}

\author{Szilard Csizmadia}
\affiliation{Institute of Planetary Research, German Aerospace Center, Rutherfordstrasse 2, 12489 Berlin, Germany}

\author{Jerome P. de Leon}
\affiliation{Department of Astronomy, University of Tokyo, 7-3-1 Hongo, Bunkyo-ku, Tokyo 113-0033, Japan}

\author{Hans Deeg}
\affiliation{Instituto de Astrof\'\i sica de Canarias, C/\,V\'\i a L\'actea s/n, 38205 La Laguna, Spain}
\affiliation{Departamento de Astrof\'isica, Universidad de La Laguna, 38206 La Laguna, Spain}

\author{Philipp Eigm\"uller}
\affiliation{Institute of Planetary Research, German Aerospace Center, Rutherfordstrasse 2, 12489 Berlin, Germany}

\author{Anders Erikson}
\affiliation{Institute of Planetary Research, German Aerospace Center, Rutherfordstrasse 2, 12489 Berlin, Germany}

\author{Mark Everett}
\affiliation{National Optical Astronomy Observatory, 950 North Cherry Avenue, Tucson, AZ 85719, USA}

\author{Malcolm Fridlund}
\affiliation{Leiden Observatory, Leiden University, 2333CA Leiden, The Netherlands}
\affiliation{Department of Space, Earth and Environment, Chalmers University of Technology, Onsala Space Observatory, 439 92 Onsala, Sweden}

\author{Akihiko Fukui}
\affiliation{Okayama Astrophysical Observatory, National Astronomical Observatory of Japan, Asakuchi, Okayama 719-0232, Japan}

\author{Eike W. Guenther}
\affiliation{Th\"uringer Landessternwarte Tautenburg, Sternwarte 5, D-07778 Tautenberg, Germany}

\author{Artie P. Hatzes}
\affiliation{Th\"uringer Landessternwarte Tautenburg, Sternwarte 5, D-07778 Tautenberg, Germany}

\author{Steve Howell}
\affiliation{Space Science and Astrobiology Division, NASA Ames Research Center, Moffett Field, CA 94035, USA}

\author{Judith Korth}
\affiliation{Rheinisches Institut f\"ur Umweltforschung an der Universit\"at zu K\"oln, Aachener Strasse 209, 50931 K\"oln, Germany}

\author{Norio Narita}
\affiliation{Department of Astronomy, University of Tokyo, 7-3-1 Hongo, Bunkyo-ku, Tokyo 113-0033, Japan}
\affiliation{Astrobiology Center, NINS, 2-21-1 Osawa, Mitaka, Tokyo 181-8588, Japan}
\affiliation{National Astronomical Observatory of Japan, NINS, 2-21-1 Osawa, Mitaka, Tokyo 181-8588, Japan}
\affiliation{Instituto de Astrof\'\i sica de Canarias, C/\,V\'\i a L\'actea s/n, 38205 La Laguna, Spain}

\author{David Nespral}
\affiliation{Instituto de Astrof\'\i sica de Canarias, C/\,V\'\i a L\'actea s/n, 38205 La Laguna, Spain}
\affiliation{Departamento de Astrof\'isica, Universidad de La Laguna, 38206 La Laguna, Spain}

\author{Grzegorz Nowak}
\affiliation{Instituto de Astrof\'\i sica de Canarias, C/\,V\'\i a L\'actea s/n, 38205 La Laguna, Spain}
\affiliation{Departamento de Astrof\'isica, Universidad de La Laguna, 38206 La Laguna, Spain}

\author{Enric Palle}
\affiliation{Instituto de Astrof\'\i sica de Canarias, C/\,V\'\i a L\'actea s/n, 38205 La Laguna, Spain}
\affiliation{Departamento de Astrof\'isica, Universidad de La Laguna, 38206 La Laguna, Spain}

\author{Martin P\"atzold}
\affiliation{Rheinisches Institut f\"ur Umweltforschung an der Universit\"at zu K\"oln, Aachener Strasse 209, 50931 K\"oln, Germany}

\author{Carina M. Persson}
\affiliation{Department of Space, Earth and Environment, Chalmers University of Technology, Onsala Space Observatory, 439 92 Onsala, Sweden}

\author{Jorge Prieto-Arranz}
\affiliation{Instituto de Astrof\'\i sica de Canarias, C/\,V\'\i a L\'actea s/n, 38205 La Laguna, Spain}
\affiliation{Departamento de Astrof\'isica, Universidad de La Laguna, 38206 La Laguna, Spain}

\author{Heike Rauer}
\affiliation{Institute of Planetary Research, German Aerospace Center, Rutherfordstrasse 2, 12489 Berlin, Germany}
\affiliation{Center for Astronomy and Astrophysics, TU Berlin, Hardenbergstr. 36, 10623 Berlin, Germany}

\author{Motohide Tamura}
\affiliation{Department of Astronomy, University of Tokyo, 7-3-1 Hongo, Bunkyo-ku, Tokyo 113-0033, Japan}
\affiliation{Astrobiology Center, NINS, 2-21-1 Osawa, Mitaka, Tokyo 181-8588, Japan}
\affiliation{National Astronomical Observatory of Japan, NINS, 2-21-1 Osawa, Mitaka, Tokyo 181-8588, Japan}

\author{Vincent Van Eylen}
\affiliation{Leiden Observatory, Leiden University, 2333CA Leiden, The Netherlands}

\author{Joshua N. Winn}
\affiliation{Department of Astrophysical Sciences, Princeton University, 4 Ivy Lane, Princeton, NJ 08544, USA}

\begin{abstract}

We present \numvalid validated planets from the 10$^\mathrm{th}$ observing campaign of the NASA \ktwo mission, as well as high resolution spectroscopy and speckle imaging follow-up observations. These \numvalid planets come from an initial set of \numtotal vetted candidates, which we subjected to a validation process incorporating pixel-level analyses, light curve analyses, observational constraints, and statistical false positive probabilities. Our validated planet sample has median values of \rp = \medrad \rearth, $P_\mathrm{orb}$ = \medper days, \teq = \medteq K, and {\em J} = \medjmag mag. Of particular interest are \nusp ultra-short period planets ($P_\mathrm{orb} \lesssim 1$\,day), \nsmall planets smaller than 2 \rearth, and \natmospheric\xspace planets with large predicted amplitude atmospheric transmission features orbiting infrared-bright stars. We also present \numcand planet candidates, most of which are likely to be real and worthy of further observations. Our validated planet sample includes \numnew new discoveries, and has enhanced the number of currently known super-Earths (\rp$\approx$ 1--2\rearth), sub-Neptunes (\rp$\approx$ 2--4\rearth), and sub-Saturns (\rp$\approx$ 4--8\rearth) orbiting bright stars ($J$ = 8--10 mag) by $\sim$4\%, $\sim$17\%, and $\sim$11\%, respectively.

\end{abstract}

\section{Introduction}

The \ktwo mission \citep{2014PASP..126..398H} is extending the \kepler legacy to a survey of the ecliptic plane, enabling the detection of transiting planets orbiting a wider range of host stars. The increased sky coverage of \ktwo has enabled the detection of planets orbiting brighter host stars, as well as a larger selection of M dwarfs \citep{2016ApJS..226....7C,2017AJ....154..207D,2018AJ....155..127H}. As a result, \ktwo is yielding a large number of promising targets for follow-up studies \citep[e.g.][]{2015ApJ...800...59V, 2015ApJ...804...10C, 2015ApJ...809...25M, 2016ApJ...827L..10V, 2015ApJ...811..102P, 2016ApJ...827L..10V, 2016ApJS..222...14V, 2016ApJ...829L...9V, 2017AJ....153..255C}. \ktwo has also discovered planets in stellar cluster environments \citep{2016AJ....152..223O,2017AJ....153..177P,2016AJ....151..112D,2016ApJ...818...46M, 2017AJ....153...64M,2017MNRAS.464..850G,2018AJ....155...10C}, including one possibly still undergoing radial contraction \citep{2016Natur.534..658D,2016AJ....152...61M}.

We present here the results of our analysis of the \ktwo photometric data collected during Campaign 10 (C10), along with a coordinated campaign of follow-up observations to better characterize the host stars and rule out false positive scenarios. Because of C10's relatively high galactic latitude, blending within the photometric apertures is less significant than for other fields, and contamination from background eclipsing binaries is low. We detect \numtotal planet candidates and validate \numvalid of them as {\it bona fide} planets using our observational constraints, \numnew of which have not previously been reported in the literature. Our sample contains a remainder of \numcand planet candidates, many of which are likely real planets.

The transit detections and follow-up observations that led to these discoveries were the result of an international collaboration called KESPRINT. Formed from the merger of two previously separate collaborations (KEST and ESPRINT), KESPRINT is focused on detecting and characterizing interesting new planet candidates from the \ktwo mission \citep[e.g.][]{2017A&A...604A..16F, 2017A&A...608A..93G, 2017AJ....154..123G, 2017AJ....154..266N, 2018MNRAS.474.5523S, 2017AJ....154..226D, 2018AJ....155..115L, 2018AJ....155..124H, 2018MNRAS.tmp.1339V}.

The rest of the paper is structured as follows. In \autoref{sec:k2} we describe our \ktwo photometry and transit search. In \autoref{sec:speckle} and \autoref{sec:spectroscopy} we describe our follow-up speckle imaging and high resolution spectroscopy of the candidates from our detection and vetting procedures. In \autoref{sec:validation} we describe our statistical validation framework and results. In \autoref{sec:discussion} we discuss particular systems of interest, and we conclude with a summary in \autoref{sec:summary}.

\section{\ktwo photometry and transit search}
\label{sec:k2}

Here we describe how we produce a list of vetted planet candidates from the pixel data telemetered from the \kepler spacecraft, as well as detailed light curve analyses. Throughout this paper we refer to stars by their nine digit EPIC IDs, and we concatenate these with two digit numbers to refer to planet candidates (ordered by orbital period).

\subsection{Photometry}

In C10, \ktwo observed a $\sim$110 square degree field near the North Galactic cap from July 06, 2016 to September 20, 2016. Long cadence (30 minute) exposures of 28,345 target stars were downlinked from the spacecraft, and the data were calibrated and subsequently made available on the Mikulski Archive for Space Telescopes\footnote{\url{https://archive.stsci.edu/k2/}} (MAST). During the beginning of the campaign, a 3.5 pixel pointing error was detected and subsequently corrected six days after the start of observations. The data during this time is of substantially lower quality than the rest of the campaign, so we discard it in our analysis. An additional data gap was the result of the failure of detector module 4, which caused the photometer to power off for 14 days.

\subsection{Systematics}

Following the loss of two of its four reaction wheels, the \kepler spacecraft has been operating as \ktwo \citep{2014PASP..126..398H}. The dominant systematic signal in \ktwo light curves is caused by the rolling motion of the spacecraft along its bore sight coupled with inter- and intra-pixel sensitivity variations. We used a method similar to that described by \citet{2014PASP..126..948V} to reduce this systematic flux variation. Our light curve production pipeline is as follows. We first downloaded the target pixel files from MAST. We laid circular apertures around the brightest pixel within the ``postage stamp'' (the set of pixels of the \kepler photometer corresponding to a given source). To obtain the centroid position of the image, we fitted a 2-D Gaussian function to the in-aperture flux distribution. We then fitted a piecewise linear function between the flux variation and the centroid motion of target. The fitted piecewise linear function was then detrended from the observed flux variation.

\subsection{Transit search}

Before searching the light curve for transits, we first removed any long-term systematic or instrumental flux variations by fitting a cubic spline to the reduced light curve from the previous section. To look for periodic transit signals, we employed the Box-Least-Squares algorithm \citep[BLS,][]{2002A&A...391..369K}. We improved the efficiency of the original BLS algorithm by using a nonlinear frequency grid that takes into account the scaling of transit duration with orbital period \citep{2014A&A...561A.138O}. We also adopted the signal detection efficiency \citep[SDE,][]{2014A&A...561A.138O} which quantifies the significance of a detection. SDE is defined by the amplitude of peak in the BLS spectrum normalized by the local standard deviation. We empirically set a threshold of SDE $>$ 6.5 for the balance between completeness and false alarm rate. In order to identify all the transiting planets in the same system, we progressively re-ran BLS after removing the transit signal detected in the previous iteration.

To search for additional transit signals which may have been missed by the transit search method described above, we used two separate pipelines: one based on the DST code \citep{2012A&A...548A..44C}, and one based on the wavelet-based filter routines {\tt VARLET} and {\tt PHALET} \citep{2016arXiv160708417G}. This helps to ensure higher detection rates, and the number of false positives is potentially reduced by utilizing multiple diagnostics. The DST code is optimized for space-based photometry and has been successfully applied to data from CoRoT and Kepler; we ran it on the light curves extracted by \citet{2014PASP..126..948V}, which are publicly available from MAST. In the wavelet-based search we first used {\tt VARLET} to remove long-term stellar variability in the light curves, and then searched for transits using a modified version of the BLS algorithm. Detected transit-like signals were then removed using {\tt PHALET}, which combines phase-folding and a wavelet basis to approximate periodic features. In similar fashion to the above approach, we iterate this process of feature detection and removal to enable the detection of multi-planet systems.

\subsection{Candidate vetting}

We performed a quick initial vetting to identify obvious false positives among the transiting signals identified in the previous section. Planetary candidates that survived the various tests were followed up with speckle imaging and reconnaissance spectra for proper statistical validation. We tested for the presence of any ``odd-even'' variations and significant secondary eclipse, both of which are likely signatures of eclipsing binaries. The odd-even effect is the variation of the eclipse depth between the primary and secondary eclipse of an eclipsing binary. If mistaken for planetary transits, the primary and secondary eclipses will be the odd and even numbered transits.

We fitted \citet{2002ApJ...580L.171M} model to the odd and even transits separately. If a systems shows odd-even variations with more than 3$\sigma$ significance, it is flagged as a false positive. We also looked for any secondary eclipse in the light curve, using the \citet{2002ApJ...580L.171M} model fit of the transits as a template for the occultation. After fitting the primary transits, we searched for secondary eclipses via an additional MCMC fitting step. We set the limb-darkening coefficients to zero and fixed all transit parameters except for two: the time of secondary eclipse and the depth of the eclipse. The resulting posterior distributions of these two parameters were then used to quantify the significance and phase of any putative secondary eclipses. For non-detections, we use the 3$\sigma$ upper limit derived from the eclipse depth posterior to set the ``maximum allowed secondary eclipse'' constraint in our {\tt vespa} analyses. If a system shows a secondary eclipse with more than 3$\sigma$ significance, we calculated the geometric albedo using the depth of secondary eclipse. The object is likely self-luminous, hence likely a false positive, if the albedo is much greater than 1.

\subsection{Stellar rotation periods}

We also measured stellar rotation periods $P_\mathrm{rot}$ from the variability in the light curves induced by starspot modulation. About half of the light curves of our candidates exhibited a lack of rotational modulation, or the \ktwo C10 time baseline was not long enough to constrain the period. For the rest, we used the autocorrelation function \citep[ACF; e.g.][]{2014ApJS..211...24M} to measure the rotational period, and we include these results in \autoref{tab:candidates} along with initial estimates of the basic transit parameters of each candidate. To help ensure the validity of these measurements, we also used the Lomb-Scargle periodogram \citep{1976Ap&SS..39..447L, 1982ApJ...263..835S} to measure the rotational periods, and the results were in good agreement.

\startlongtable
\begin{deluxetable}{lllllllc}
\tabletypesize{\tiny}
\tablecaption{Candidate planets detected in \ktwo C10. $Kp$ denotes magnitude in the \kepler bandpass. \label{tab:candidates}}
\tablehead{EPIC & $Kp$ & $P_\mathrm{orb}$ & $T_0$ & $T_{14}$ &  Depth & SDE & $P_\mathrm{rot}$ \\
           & [mag] & [days] & [BKJD] & [hours] & & & [days] }
\startdata
\input{tab_candidates.tex}
\enddata
\end{deluxetable}

\subsection{Transit modeling}

We used the orbital period, mid-transit time, transit depth, and transit duration identified by BLS as the starting points for more detailed transit modeling. The transit light curve was generated by the {\tt Python} package {\tt batman} \citep{2015PASP..127.1161K}. To reduce the data volume, we only use the light curve in a 3$\times T_{14}$ window centered on the mid-transit times. We first tested if any of the systems showed strong transit timing variations (TTVs). We used the Python interface to the Levenberg-Marquardt non-linear least squares algorithm {\tt lmfit} \citep{newville_2014_11813} to find the best-fit model of the phase-folded transit, and then fit this template to each transit separately to identify individual transit times of each candidate. Since none of the system presented in this work showed significant TTVs within the \ktwo C10 observations, we assumed linear ephemerides in subsequent analyses.

The transit parameters in our linear ephemeris model include the orbital period $P_\mathrm{orb}$, the mid-transit time $T_0$, the planet-to-star radius ratio \rp/\rstar, the scaled orbital distance $a$/\rstar, the impact parameter $b\equiv a\cos i/R_\star$, and the transformed quadratic limb-darkening coefficients $q_1$ and $q_2$. Instead of fixing the parameters of the quadratic limb-darkening law to theoretical values based on stellar models, in this work we opt to allow these parameters to vary, as this allows for error propagation from stellar uncertainties. We utilize the available stellar parameters and their uncertainties to impose Gaussian priors on the limb-darkening coefficients (i.e. in the non-transformed parameter space, $u_1$ and $u_2$). To determine the location and width of these priors, we used a Monte Carlo method to sample the stellar parameters of each candidate host star (\teff, \logg, and \feh), and then used these to derive distributions of $u_1$ and $u_2$ from an interpolated grid based on the limb-darkening coefficients for the Kepler bandpass tabulated by \citet{2012yCat..35460014C}. We used the median and standard deviation of these distributions to define the Gaussian limb-darkening priors, and used uniform priors for all other parameters. Depending on the uncertainty in the stellar parameters, the limb-darkening priors determined in this way have typical widths of $\sim$10\%, which is comparable to the uncertainty in the models used to predict them \citep[e.g.][]{2013A&A...549A...9C, 2013A&A...560A.112M}. In addition, when the stars are active we do not expect agreement between theoretical and observed limb darkening because the tabulated theoretical values do not take into account the effects of stellar spots and faculae \citep{2013A&A...549A...9C}. To account for the 30\,min integration time of long cadence \ktwo photometry, we used the built-in feature of {\tt batman} to super-sample the model light curve by a factor of 16 before averaging every 3\,min window \citep{2010MNRAS.408.1758K}.

\begin{figure*}
    \centering
    \includegraphics[width=0.95\textwidth,trim={0.5cm 0 0.5cm 0}]{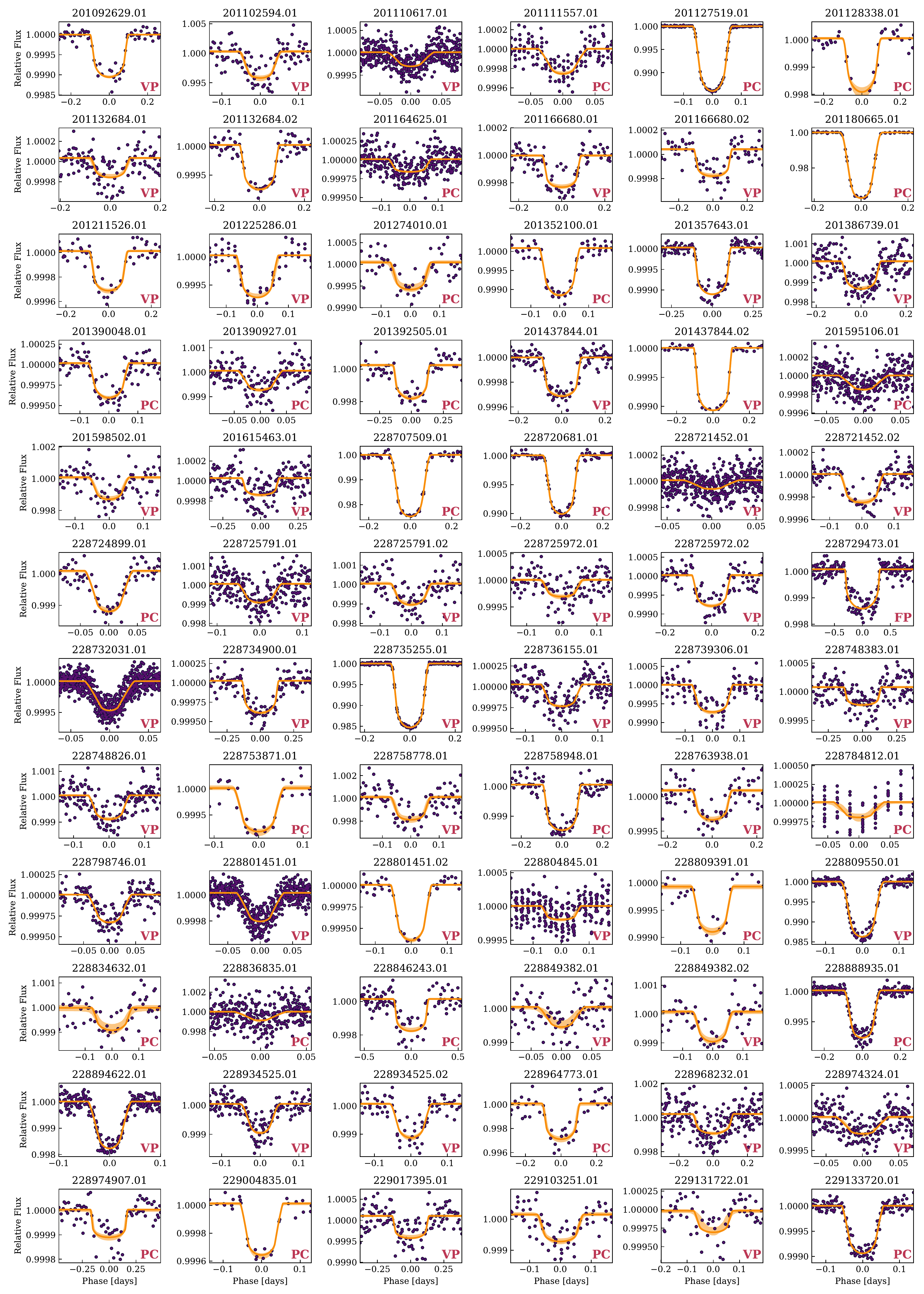}
    \caption{Phase-folded transits (purple), with the best-fit transit model and 1$\sigma$ credible region overplotted (orange). Candidate dispositions are displayed in the lower-right corners (see \autoref{sec:validation}).}
    \label{fig:plots}
\end{figure*}

We adopted a Gaussian likelihood function, and found the maximum likelihood solution using {\tt scipy.optimize} \citep{scipy}. We then sampled the joint posterior distribution using {\tt emcee} \citep{emcee}, a Python implementation of the affine-invariant Markov Chain Monte Carlo ensemble sampler \citep{2010CAMCS...5...65G}. We assumed the errors to be Gaussian, independent, and identically distributed, and thus described by a single parameter. In the maximum likelihood fits, we fixed the value of this parameter to the standard deviation of the out of transit flux, and during MCMC we fit for this value as a free parameter. We launched 100 walkers in the vicinity of the maximum likelihood solution and ran the sampler for 5000 steps, discarding the first 1000 as ``burn-in.'' To ensure that the resultant marginalized posterior distributions consisted of 1000's of independent samples (enough for negligible sampling error) we computed the autocorrelation time of each parameter, and visual inspection revealed the posteriors to be smooth and unimodal. We summarize the transit parameter posterior distributions in \autoref{tab:params} using the 16$^{th}$, 50$^{th}$, and 84$^{th}$ percentiles, and we use the posterior samples to compute other quantities of interest throughout this work (i.e. \rp, \teq). The phase-folded light curves of the candidates are shown in \autoref{fig:plots}, with best-fitting transit model and 1$\sigma$ (68\%) credible region over-plotted.

\section{Speckle imaging}
\label{sec:speckle}

We observed candidate host stars with the NASA Exoplanet Star and Speckle Imager (NESSI) on the 3.5-m WIYN telescope at the Kitt Peak National Observatory. NESSI is a new instrument that uses high-speed electron-multiplying CCDs (EMCCDs) to capture sequences of 40 ms exposures simultaneously in two bands (\citet{2016SPIE.9907E..2RS}, Scott et al., in prep.). Data were collected following the procedures described by \citet{2011AJ....142...19H}. We conducted all observations in two bands simultaneously: a `blue' band centered at 562nm with a width of 44nm, and a `red' band centered at 832nm with a width of 40nm. The pixel scales of the `blue' and `red' EMCCDs are 0.0175649\arcsec\, and 0.0181887\arcsec\, per pixel, respectively. We make all of our speckle imaging data publicly available via the community portal ExoFOP\footnote{\url{https://exofop.ipac.caltech.edu}}. We list the individual NESSI data products used in this work in \autoref{tab:speckle_obs}.

Speckle imaging data were reduced following the procedures described by \citet{2011AJ....142...19H}, resulting in diffraction limited $4.6\arcsec\times4.6\arcsec$ reconstructed images ($256\times256$ pixels) of each target star. The methodology has been described in detail in previous works \citep[e.g.][]{2009AJ....137.5057H, 2012AJ....144..165H, 2017AJ....153..212H}, but we provide a brief review here for convenience.

First, the autocorrelation function of each 40 ms exposure is summed and Fourier transformed, resulting in the average spatial frequency power spectrum. The speckle transfer function is then deconvolved by dividing the target's power spectrum by that of the corresponding point source calibrator, yielding the square of the modulus estimate of the target's Fourier transform. The phase information can then be recovered from bispectral analysis, as first described by \citet{1983ApOpt..22.4028L}. This is accomplished by computing the Fourier transform of the summed triple correlation function of the exposures, which in combination with the modulus estimate yields the complex Fourier transform of the target. This is then filtered with a low-pass 2-d Gaussian before being inverse transformed, yielding the reconstructed image.

\begin{figure}
    \centering
    \includegraphics[width=0.49\textwidth,trim={0 0 0 0}]{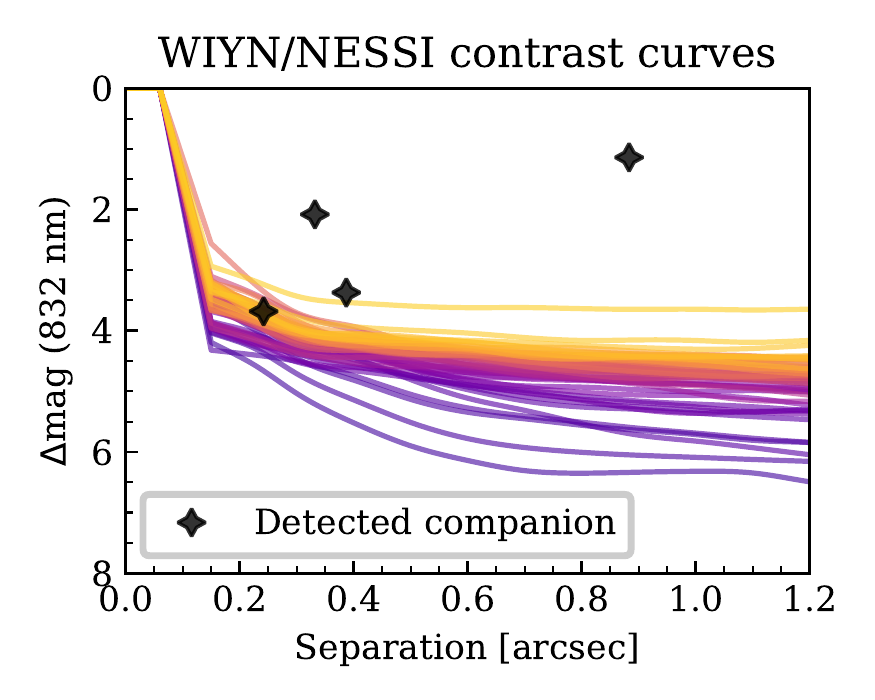}
    \caption{Contrast curves and detected companions}
    \label{fig:cc}
\end{figure}

We extract background sensitivity limits from the reconstructed images by computing the mean and standard deviation of a series of concentric annuli centered on the target star, as described by \citet{2011AJ....142...19H}. We then compute contrast curves by fitting a cubic spline to the kernel-smoothed 5$\sigma$ sensitivity limits, expressed as a magnitude difference relative to the target star as a function of radius. For stars of moderate brightness ($V = 10-12$ mag) we typically achieve contrasts of $\sim$ 4 magnitudes at 0.2\arcsec. See \autoref{fig:cc} for a plot showing all of the contrast curves obtained in this work. We detect 4 candidate host stars with secondaries, see \autoref{tab:companions}.

\begin{deluxetable}{lllll}
\tabletypesize{\scriptsize}
\tablecaption{Stars with detected companions. All detections made in the 832nm band. \label{tab:companions}}
\tablehead{ EPIC &    $\Delta$arcsec &  $\Delta$mag &    $\theta$ [deg. E of N] & Note}
\startdata
 201352100 &  0.387 &  3.37 &  312.054 &     a \\
 201390927 &  0.883 &  1.14 &  341.286 &     a \\
 201392505 &  0.242 &  3.68 &   42.491 &     b \\
 228964773 &  0.332 &  2.08 &   43.499 &     b \\
\enddata
\tablecomments{a: The quadrant of the position angle is ambiguous, meaning it could be off by exactly 180 degrees. b: The binary model fit is of poor quality, so uncertainty may be larger than typical.}
\end{deluxetable}

\begin{figure}
    \centering
    \includegraphics[width=0.5\textwidth,trim={2cm 1.5cm 1cm 0.5cm}]{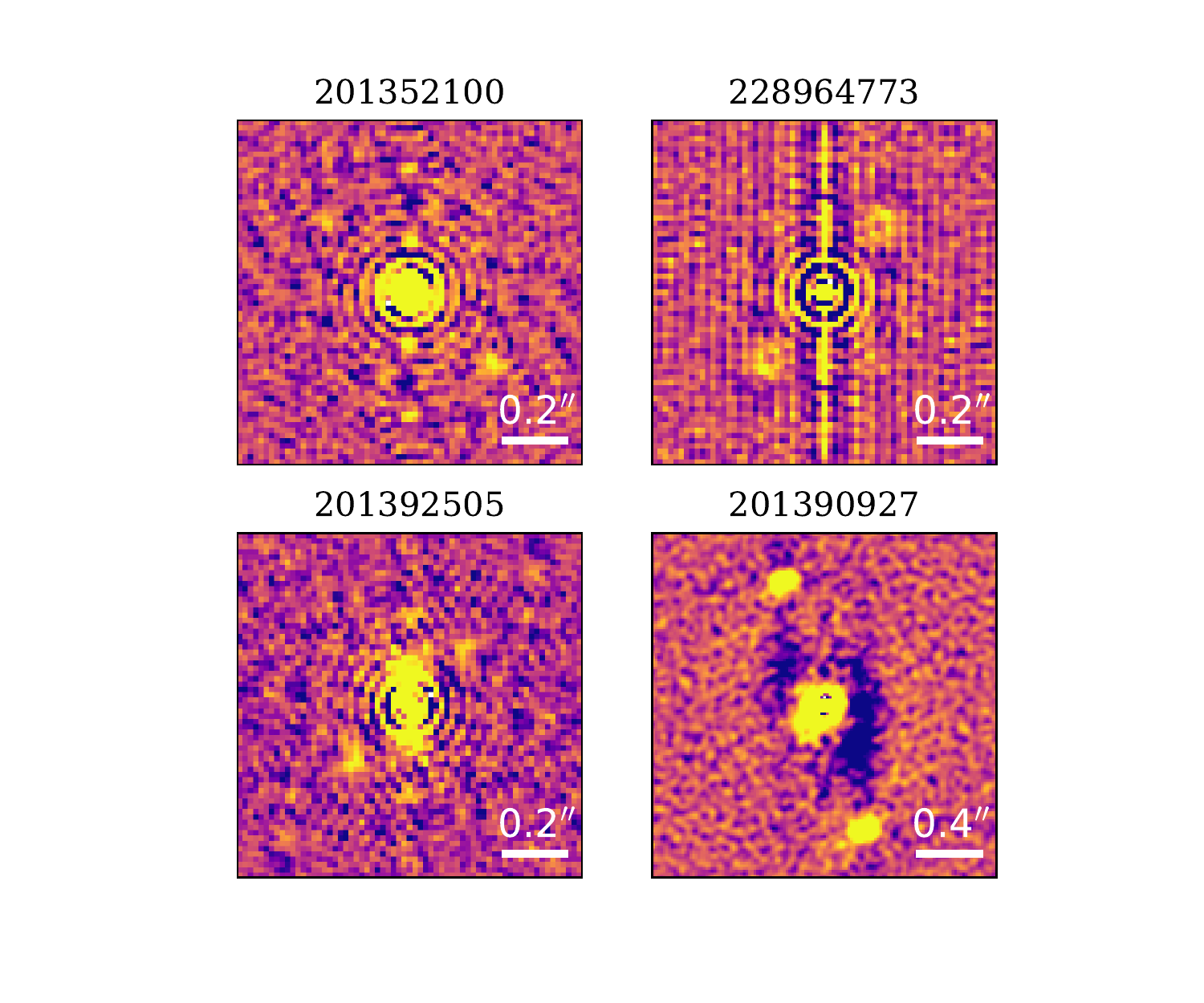}
    \caption{Reconstructed 832nm images of stars with detected companions.}
    \label{fig:companions}
\end{figure}

\section{High resolution spectroscopy}
\label{sec:spectroscopy}

\subsection{McDonald/Tull}
\label{sec:tull}

Most of the high resolution spectra presented in this paper were obtained with the Tull Coud\'{e} cross-dispersed echelle spectrograph \citep{1995PASP..107..251T} at the Harlan J. Smith 2.7m telescope at McDonald Observatory. Observations were conducted with the 1.2$\times$8.2\arcsec\, slit, yielding a resolving power of $R \sim 60,000$. The spectra cover 375-1020\,nm, with increasingly larger inter-order gaps long-ward of 570\,nm. For each target star, we obtained three successive short exposures in order to allow removal of energetic particle hits on the CCD detector. We used an exposure meter to obtain an accurate flux-weighted barycentric correction and to give an exposure length that resulted in a signal/noise ratio of about 30 per pixel.  Bracketing exposures of a Th-Ar hollow cathode
lamp were obtained in order to generate a wavelength calibration and to remove spectrograph drifts.   This enabled calculation of absolute radial velocities from the spectra.  The raw data were processed using IRAF routines to remove the bias level, inter-order scattered light, and pixel-to-pixel (``flat field'') CCD sensitivity variations.  We traced the apertures for each spectral order and used an optimal extraction algorithm to obtain the detected stellar flux as a function of wavelength.

We computed stellar parameters from our reconnaissance Tull spectra using {\it Kea} \citep{2016PASP..128i4502E}. In brief, we used standard IRAF routines to perform flat fielding, bias subtraction, and order extraction, and we used a blaze function determined from high SNR flat field exposures to correct for curvature induced by the blaze. {\it Kea} uses a large grid of synthetic model stellar spectra to compute stellar effective temperatures, surface gravities, and metallicities.
See \autoref{tab:stellar} for the stellar parameters used in this work. From a comparison with higher SNR spectra obtained with Keck/HIRES we found typical uncertainties of 100 K in \teff, 0.12 dex in \feh, and 0.18 dex in \logg. For a detailed description of Kea see \citet{2016PASP..128i4502E}.

\subsection{NOT/FIES}

We also used the FIbre-fed \'Echelle Spectrograph \citep[FIES;][]{Frandsen1999,Telting2014} on the
2.56-m Nordic Optical Telescope (NOT) of Roque de los Muchachos Observatory (La Palma, Spain) to collect high-resolution ($R\,\approx\,67\,000$) spectra of four C10 candidate host stars: 228729473, 228735255 (K2-140; \citet{2018MNRAS.475.1809G}, Korth et al., submitted to MNRAS), 201127519, and 228732031 \citep[K2-131;][]{2017AJ....154..226D}. The observations were carried out between February 15 to May 23, 2017 UTC, within observing programs 54-027, 55-019, and 55-202. We followed the same strategy as in \citet{Gandolfi2013} and traced the RV drift of the instrument by bracketing the science exposures with 90-sec ThAr spectra. We reduced the data using standard IRAF routines and extracted the RVs via multi-order cross-correlations using different RV standard stars observed with the same instrument.

\subsection{TNG/HARPS-N}

We observed the stars 228801451, 228732031 \citep[K2-131;][]{2017AJ....154..226D}, 201595106, and 201437844 \citep[HD\,106315;][]{2017AJ....153..255C,2017AJ....153..256R} with the HARPS-N spectrograph \citep[$R\,\approx$\,115000;][]{Cosentino2012} mounted at the 3.58\,m Telescopio Nazionale Galileo (TNG) of Roque de los Muchachos Observatory (La Palma, Spain). The observations were performed in January 2017 as part of observing programs A34TAC\_10 and A34TAC\_44. We reduced the data using the dedicated off-line pipeline and extracted the RVs by cross-correlating the \'echelle spectra with a G2 numerical mask. The HARPS-N data of 228732031 have been published by our team in \citet{2017AJ....154..226D}. We refer the reader to that paper for a detailed description and analysis of the data. We list the results of our analysis of these spectra in \autoref{tab:harps}.

\subsection{Stellar properties}
\label{sec:stellar}

We obtained spectra for 27 candidate host stars in this work, from which we derived \teff, \logg, \feh, and \vsini, as described in \autoref{sec:tull}. We augment this set of spectroscopic stellar parameters with values from the literature for an additional 14 candidate host stars (\citealt{2017AJ....153..256R,2018AJ....155..127H,2018AJ....155..136M}). To maximize both the quality and uniformity of the final set of stellar parameters we use in this work, we adopted the following strategy. First, we gathered 2MASS $JHK$ photometry and Gaia DR2 parallaxes for all stars; 2MASS photometry is available in the EPIC, and we cross-matched to Gaia DR2 using both position and optical magnitude agreement ($Kp$ and Gaia $G$ band). We then used the {\tt isochrones} \citep{2015ascl.soft03010M} interface to the Dartmouth stellar model grid \citep{2008ApJS..178...89D} to estimate stellar parameters and their uncertainties using the {\tt MultiNest} sampling algorithm \citep{2013arXiv1306.2144F}. For those stars with parameters from spectroscopic analyses, we imposed Gaussian priors on \teff, \logg, and \feh, with mean and standard deviation set by the spectroscopically derived values and their uncertainties. We also ran the same analysis without including parallax, as a check on the quality of the parameters derived in this manner without any distance information; unsurprisingly, we found that including parallax yielded the biggest improvement for stars lacking spectroscopy. This is perhaps most important for the M dwarfs in our sample, which suffer from systematically underestimated radii in the EPIC \citep[see e.g.][]{2017AJ....154..207D}.

As an additional quality check, we also performed spectral analyses for the targets 201127519, 201437844, 201595106, and 228801451, using spectra from FIES and HARPS-N and {\tt SpecMatch-emp} \citep{2017ApJ...836...77Y}. {\tt SpecMatch-emp} fits the input spectra to hundreds of library template spectra collected by the California Planet Search, and the stellar parameters (\teff, \rstar, and \feh) are estimated based on the interpolation of the parameters for best-matched
library stars. Among them 201127519, 201595106, and 228801451 were also
observed with the Tull spectrograph, and the resulting parameters by {\tt SpecMatch-emp} are in agreement
within $\sim 1.5\sigma$ with those estimated from the Tull spectra by the {\it Kea} code.
For HD\,106315, we obtained \teff = $6326\pm110$ K, \rstar = $1.86\pm0.30$ \rsun,
and \feh = $-0.20\pm0.08$. While \teff and \feh agrees within
1$\sigma$ with the literature values \citep{2017AJ....153..256R, 2017AJ....153..255C},
\rstar exhibits a moderate disagreement with that in the literature
\citep[\rstar = $1.281_{-0.058}^{+0.051}$ \rsun][]{2017AJ....153..256R}.
This is probably due to the small number of library stars in {\tt SpecMatch-emp}
in the region with \teff $>6300$ K, but this disagreement does not have
any impact on our results.

\section{Planet validation}
\label{sec:validation}

\begin{figure*}
    \centering
    \subfloat{\includegraphics[clip,trim={0 0 0 0},width=0.5\textwidth]{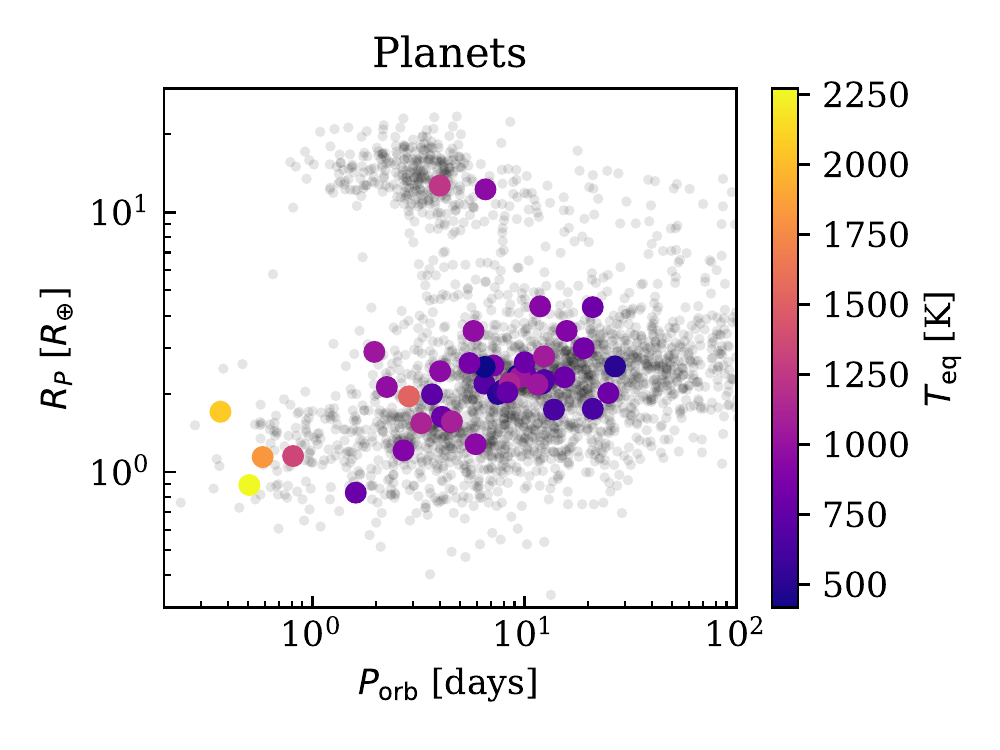}}
    \subfloat{\includegraphics[clip,trim={0 0 0 0},width=0.5\textwidth]{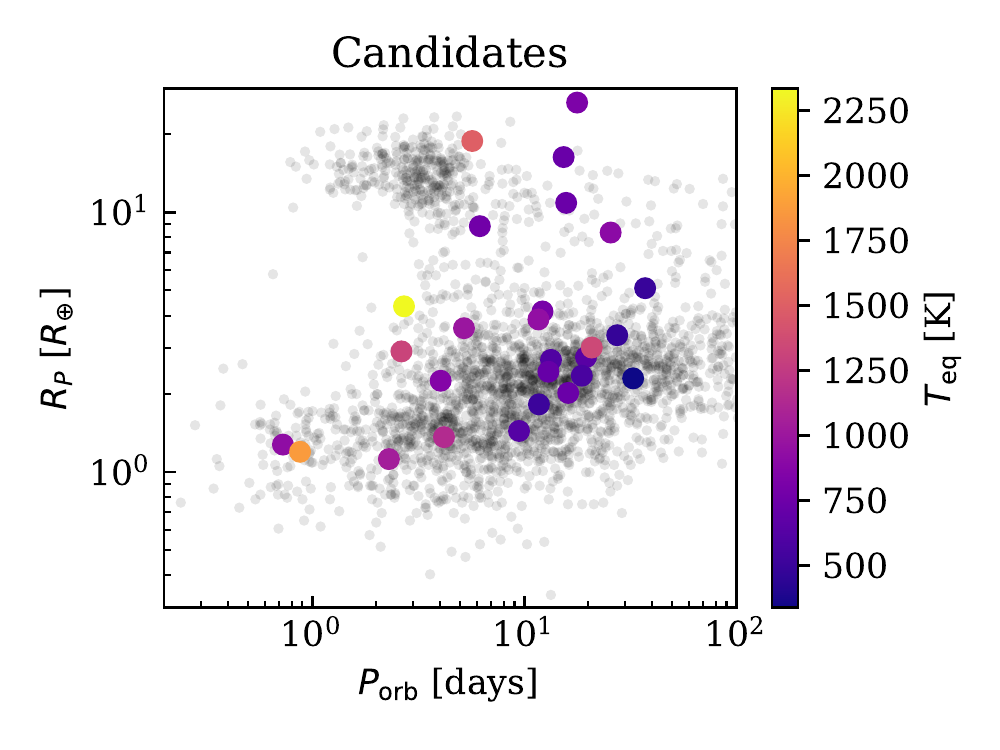}}
    \caption{Validated (left) and candidate (right) planets from C10 against the background of previously confirmed or validated planets, colored by their equilibrium temperature (assuming a Bond albedo of 0.3).}
    \label{fig:planets}
\end{figure*}

\subsection{Statistical framework}

We use the open source {\tt vespa} software package \citep{2012ApJ...761....6M,2015ascl.soft03011M} to compute the false positive probabilities (FPPs) of each planet candidate. {\tt vespa} uses the {\tt TRILEGAL} Galaxy model \citep{2005A&A...436..895G} to compute the posterior probabilities of both planetary and non-planetary scenarios given the observational constraints, and considers false positive scenarios involving simple eclipsing binaries, blended background eclipsing binaries, and hierarchical triple systems. {\tt vespa} models the physical properties of the host star, taking into account any available broadband photometry and spectroscopic stellar parameters, and compares a large number of simulated scenarios to the observed phase-folded light curve. Both the size of the photometric aperture and contrast curve constraints are accounted for in the calculations, as well as any other observational constraints such as the maximum depth of secondary eclipses allowed by the data. We adopt a fiducial validation criterion of FPP $<$ 0.01, which is reasonably conservative and also consistent with the literature \citep[e.g.][]{2015ApJ...809...25M,2016ApJS..226....7C,2016ApJ...822...86M}. {\tt vespa} utilizes the contrast curves derived from the observations listed in \autoref{tab:speckle_obs} and described in \autoref{sec:speckle}. To minimize the possibility of errors in the {\tt vespa} calculations induced by zero-point offsets or underestimated uncertainties in broadband photometry, we opt to use only the well-calibrated 2MASS $JHK$ magnitudes and their uncertainties, taken from the EPIC, in addition to the \kepler band magnitude required by {\tt vespa}. The stellar parameters used as input to {\tt vespa} are identical to those used in our uniform {\tt isochrones} analysis (see \autoref{sec:stellar}). In addition to stellar parameters, {\tt vespa} utilizes basic system properties (i.e. RA, Dec, $P_\mathrm{orb}$, \rp/\rstar), as well as contrast curves (see \autoref{sec:speckle}) and constraints on secondary eclipse depth and maximum exclusion radii (see \autoref{tab:vespa_constraints}). We tabulate candidate parameters along with their FPPs and final dispositions in \autoref{tab:params}, and the full {\tt vespa} likelihoods are listed in \autoref{tab:vespa}. We denote final dispositions as follows: ``VP'' = validated planet; ``PC'' = planet candidate; ``FP'' = false positive.

All of the candidates we detect in multi-planet systems meet the fiducial validation criterion of FPP $<$ 1\%. However, FPPs computed with {\tt vespa} treat only the individual planet candidates in isolation, and thus do not take into account any multiplicity in each system. Stars with multiple transiting planet candidates have been shown to exhibit a lower false positive rate by an order of magnitude \citep{2011ApJS..197....8L,2012ApJ...750..112L,2014ApJ...784...44L}. For this reason we apply a ``multiplicity boost'' factor to the planet probability appropriate for each candidate in a multi-planet system. \citet{2012ApJ...750..112L} estimated a multiplicity boost factor of 25 for systems containing 2 planet candidates in the \kepler field, and we apply the same factor in this work. To check that this factor is appropriate for \ktwo C10, we follow \citet{2016ApJ...827...78S} and utilize equations (2) and (4) of \citet{2012ApJ...750..112L} to estimate the sample purity $P$ from the integrated FPP of our sample and the number of planet candidates we detect (\numtotal). This estimate of $P$ is quite high, perhaps due to a lack of contamination from background stars due to the high galactic latitude of the field, or due to our team's vetting procedures. The fraction of detected planet candidates in multi-systems (\nummulti/\numtotal) in conjunction with the high sample purity yields a multiplicity boost which is significantly higher than the factor of 25 estimated by \citet{2012ApJ...750..112L} for the \kepler field. Although the true value is likely to be higher, we conservatively apply only a factor of 25, consistent with \citet{2012ApJ...750..112L}, and the FPPs in \autoref{tab:params} reflect this accordingly.

\subsection{Stellar companions}
\label{sec:companions}

To ensure that the FPPs computed by {\tt vespa} are reliable, we take into account the presence of any nearby stars detected in speckle or archival imaging. \autoref{tab:companions} lists the nearby stars we detected via speckle imagine, along with their separations and delta-magnitudes relative to the primary stars. \autoref{fig:companions} shows the reconstructed speckle images for these stars, and \autoref{fig:cc} shows these detections relative to the ensemble of contrast curves from all of our speckle images. \autoref{tab:epic-companions} lists those stars found in the EPIC to be near and bright enough to be the source of the observed transit signals.

\subsubsection{Companions detected in high resolution imaging}

On the nights of 2017-03-15, 2017-03-17, and 2017-03-18 we acquired speckle imaging of the stars 201352100, 201390927, 201392505, and 228964773 (see \autoref{tab:speckle_obs}).
We detected companions in the reconstructed images (see \autoref{fig:companions}), so we assessed the possibility that the transit signal might not originate from the primary stars. We used the following relation between the observed transit depth $\delta'$ and the true transit depth $\delta$ in the presence of dilution from a companion $\Delta m$ magnitudes fainter than the primary star:
\begin{equation}
  \delta' = \frac{\delta}{1 + 10^{0.4\Delta m}}
\end{equation}
Assuming a maximum eclipse depth of 100\% (i.e. a brown dwarf --- M dwarf binary) we can potentially rule out the secondary star as the source of the observed signal. For shallower transits the maximum allowed dilution from the primary is larger, and therefore even a relatively faint secondary source cannot be ruled out as the host. For each of these four of these candidates, the secondary source is bright enough (given the observed transit depth) that we cannot rule out the possibility they are the source of the signal (see \autoref{tab:companions}). For this reason, we do not validate any of these candidates as planets, as we do not know the true source of the signal (and therefore the true planet size), even though they all have low FPPs.

\subsubsection{Companions in the EPIC}

In addition to analyzing the scenarios involving companions detected in high resolution speckle imaging, we also performed a search of the EPIC for any additional stars within the photometric apertures which could be the source of the observed signals. Most of these queries yielded no stars within the aperture other than the primary, but there were some cases in which the query yielded a star bright enough to be the source of the observed transit signal; we list these cases in \autoref{tab:epic-companions}. Despite their low FPPs, we do not validate these candidates because we do not know which star is the true host. As we expect most of these candidates to be genuine planets, they present good validation opportunities via higher angular resolution follow-up transit observations, either from the ground or from space (i.e. with \spitzer or \cheops).

\begin{deluxetable}{lccc}
\tabletypesize{\scriptsize}
\tablecaption{EPIC sources within the photometric apertures which are bright enough to produce the observed transit-like signals. \label{tab:epic-companions}}
\tablehead{EPIC & Contaminant & $\rho$ [arcsec] & $\Delta Kp$ [mag] \\}
\startdata
\input{tab_epic_companions.tex}
\enddata
\end{deluxetable}

\subsubsection{Archival imaging}

\begin{figure}
    \centering
    \includegraphics[clip,trim={0cm 0 0cm 0},width=0.48\textwidth]{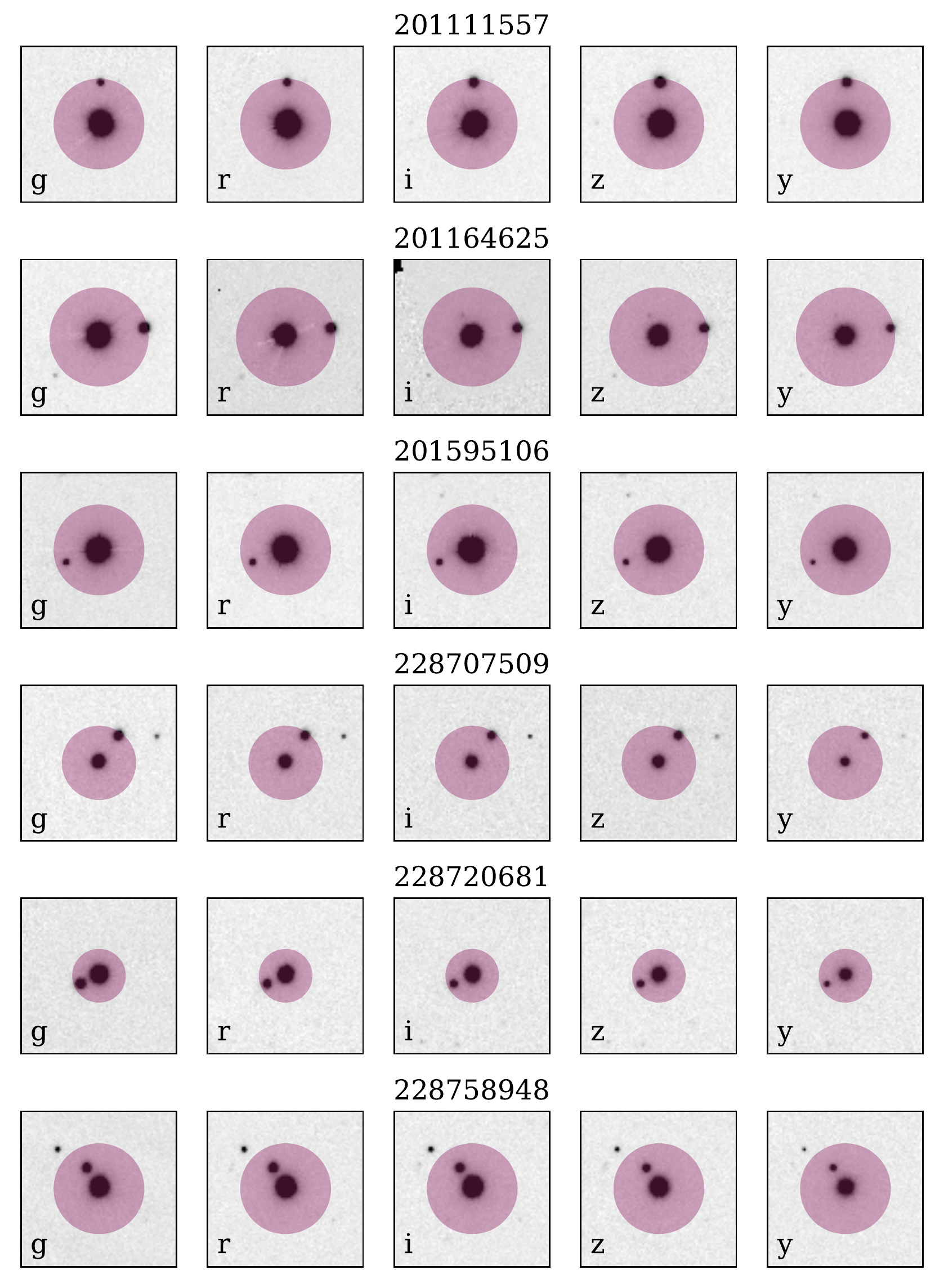}
    \caption{Archival {\it grizy} imaging from Pan-STARRS-1. Shown here are candidate planet hosts with nearby bright stars within the \ktwo apertures (represented by circular shaded regions). Assuming a maximum eclipse depth of 100\%, the observed transit-like signal could potentially be reproduced by scenarios in which the signal is actually a faint eclipsing binary diluted by the flux from the brighter primary star. We note, however, that such scenarios would sometimes result in more ``V-shaped'' transits than what we observe.}
    \label{fig:panstarrs}
\end{figure}

As a check on the accuracy of the sources comprising the EPIC, we also queried 1'$\times$1' Pan-STARRS-1\footnote{Data release 1, dated December 19, 2016, available at http://ps1images.stsci.edu/cgi-bin/ps1cutouts} {\it grizy} images centered at the position of each candidate host star. We found good agreement with the catalog query: nearby stars found by the catalog query were clearly visible in the images, and no nearby bright sources were seen in the images that were not previously found by the catalog query. We show these images in \autoref{fig:panstarrs}, with overplotted circular regions illustrating the size and location of the apertures used to extract photometry from the \ktwo pixel data.

\subsection{Multi-aperture light curve analysis}

In light of several recent cases of contamination from false positives in statistically validated planet samples \citep{2017ApJ...847L..18S, 2017A&A...606A..75C}, we also scrutinized our candidates at the pixel level. To do so, we extracted light curves from different sized apertures and looked for signs of a dependence of transit depth on aperture radius. In some cases, these light curves are too noisy to draw conclusions from, as they are extracted from ``non-optimal'' apertures. However, this analysis is especially important when there are widely separated neighboring stars (i.e. several \kepler pixels away) that still contribute flux to the \ktwo apertures, in which case it may be possible to determine the origin of the transit-like signal by this method. Based on these analyses we found that the transit signal associated with the candidate 201164625.01 most likely originates from the neighboring star, 201164669 (see \autoref{tab:epic-companions} and \autoref{fig:panstarrs}). We also detected suspicious transit depth behavior in the light curves of 201392505.01 and 228964773.01, both of which have nearby companions detected in speckle imaging. Intriguingly, these companions are well within a \kepler pixel of the target star, so even the smallest aperture possible (one \kepler pixel) should contain light from both the primary and secondary stars. This result may indicate the presence of another (undetected) star further away, and suggests that such multi-aperture analyses should be useful for ranking the quality of candidates when high resolution imaging is unavailable.

\subsection{Transit SNR}

As a final step in the validation process, we compute the transit SNR for each candidate in order to enforce a minimum transit quality standard for all planets in the validated sample. We compute the transit SNR using the simple approximation that the signal scales with the transit depth and the square root of the number of transits \citep[e.g.][]{2017arXiv170508891B}. We estimate the noise by computing the standard deviation of the out-of-transit photometry used in our light curve fits and scaling it from the \ktwo observing cadence to the transit duration of each candidate. We find median SNR values of 17.1 and 17.6 for the validated and candidate samples, respectively. The slightly lower SNR of the validated sample is likely attributable to the fact that candidates with higher FPPs are typically larger and have correspondingly deeper transits, whereas the vast majority of our validated planets are sub-Neptunes (see \autoref{fig:planets}). Our validated sample consists of planets with SNR $>$ 10, with the exception of K2-254\,b and K2-247\,c, which have SNR values of 6.7 and 8.9, respectively. However, these are both in multi-planet systems, which increases our confidence in the veracity of the transit signals. We argue that candidates with relatively low SNR found in systems with multiple validated candidates need not be regarded with as much suspicion as similarly low SNR candidates in single-candidate systems; this is related to, but more qualitative than, the ``multi-boost'' argument of \citet{2012ApJ...750..112L}. Indeed, many interesting planets with low SNR likely remain to be found in both the \kepler and \ktwo data \citep[e.g.][]{2018AJ....155...94S}.

\subsection{Pipeline comparison}

To check the quality of our light curves and provide an additional layer of confidence in our candidates, we performed a parallel analysis using light curves from an independent \ktwo pipeline. We first downloaded the light curves of \citet{2014PASP..126..948V} from MAST for all the targets listed in \autoref{tab:candidates}, then detrended the light curves by fitting a second order polynomial to the out-of-transit data using \texttt{exotrending} \citep{2017ascl.soft06001B}. To explore the transit model parameter space with MCMC, we used \texttt{pyaneti} \citep{2017ascl.soft07003B} to fit the detrended light curves with uniform priors for all parameters; more description of the \texttt{pyaneti} MCMC evolution and parameter estimation can be found in \citet{2017arXiv171102097B} and \citet{2017AJ....154..123G}. For the majority of candidates, the main transit parameters of interest ($P_\mathrm{orb}$, \rp/\rstar, $b$, and $a$/\rstar) are consistent within 1$\sigma$ between our two independent analyses, although there are some cases in which marginally significant differences were found. These differences are likely to be the result of different handling of the \ktwo systematics and/or the stellar variability in the light curves. The overall good agreement between these two independently-derived sets of transit parameters provides an additional layer of confidence in the quality of the candidates. The results of this comparison are listed in \autoref{tab:comparison}.

\section{Discussion}
\label{sec:discussion}

\subsection{Validated planets}

We validate \numvalid planets out of our sample of \numtotal candidates, and tabulate the FPPs along with parameter estimates of interest in \autoref{tab:params}. Of the \numvalid validated planets we report here, 20 of them have been previously statistically validated or confirmed: 201598502.01, 228934525.01, and 228934525.02 \citep[K2-153\,b, K2-154\,bc;][]{2018AJ....155..127H}; 228735255.01 (K2-140\,b; \citet{2018MNRAS.475.1809G}, Korth et al., submitted to MNRAS); 201437844.01 and 201437844.02 \citep[HD\,106315\,bc;][]{2017AJ....153..255C, 2017AJ....153..256R}; 228732031.01 \citep[K2-131\,b;][]{2017AJ....154..226D}; and 13 others were recently validated by \citet{2018AJ....155..136M}. In the left panel of \autoref{fig:planets} we plot the planetary radii, orbital periods, and equilibrium temperatures of the validated planets in the sample.

\begin{figure}
    \centering
    \includegraphics[clip,trim={0 0 0 0},width=0.5\textwidth]{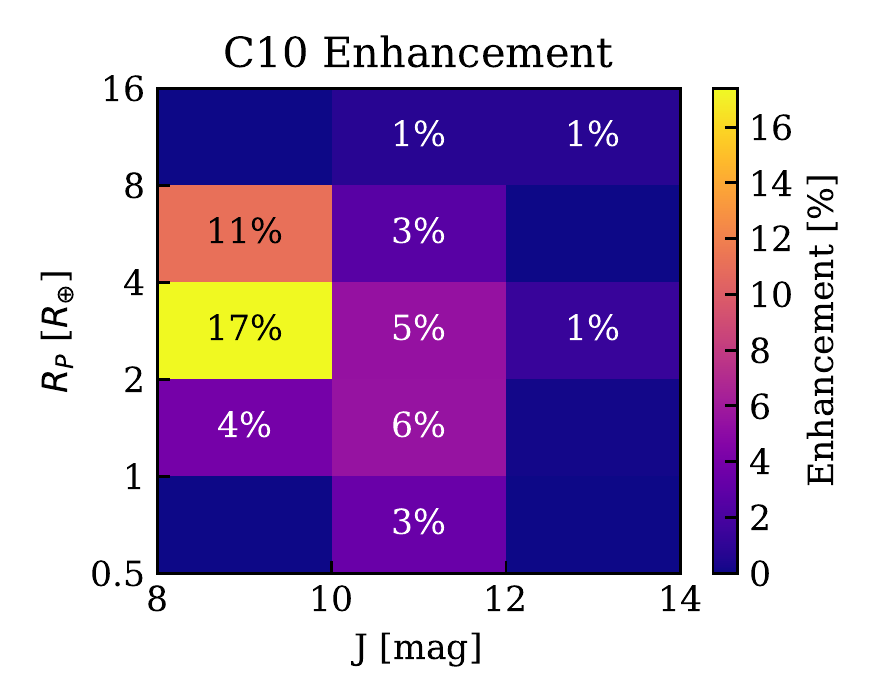}
    \caption{The fractional enhancement to the population of previously validated or confirmed planets from our sample of \numvalid validated C10 planets.}
    \label{fig:enhancement}
\end{figure}

We investigated the impact of these new planets to the population of currently known planets by querying the NASA Exoplanet Archive\footnote{\url{https://exoplanetarchive.ipac.caltech.edu/}} \citep{2013PASP..125..989A}. We computed the fractional enhancement to the known population due to the \numvalid planets as a function of planet size and host star brightness (see \autoref{fig:enhancement}). As of June 12, 2018, the populations of super-Earths (\rp$\approx$ 1--2\rearth), sub-Neptunes (\rp$\approx$ 2--4\rearth), and sub-Saturns (\rp$\approx$ 4--8\rearth) orbiting bright stars ($J$ = 8--10 mag) are enhanced by $\sim$4\%, $\sim$17\%, and $\sim$11\%, respectively. Because of the brightness of the host stars, many of these planets are ideal for detailed characterization studies via precision Doppler and transmission spectroscopy, which we discuss in greater detail in \autoref{sec:characterization}.

\subsection{Candidates}
\label{sec:candidates}

Out of the 72 planet candidates we present here, \numcand are not validated. Most cannot be validated due to the FPP being above our fiducial validation criterion of 1\% or the presence of a contaminating star within the photometric aperture. See \autoref{tab:vespa} for the likelihoods of various false positive scenarios and the planet scenario, as computed by {\tt vespa}. There are several candidates which we do not validate for other reasons, which we discuss below. In the right panel of \autoref{fig:planets} we plot the planetary radii, orbital periods, and equilibrium temperatures of the non-validated candidates.

The candidate 228729473.01 exhibits a long transit duration, and subsequent spectroscopic analyses revealed large RV variations which are consistent with the candidate being a false positive involving an M dwarf eclipsing a sub-giant, see Csizmadia et al (in prep.) for more details. The light curve of 229133720.01 exhibits low levels of variability in phase with the transit signal, which could be due to ellipsoidal variations; thus we do not validate the candidate in spite of its low FPP. Although 201390048.01 was recently validated \citep[K2-162\,b;][]{2018AJ....155..136M}, we found marginal evidence of odd-even variations in the light curve of this candidate, which could be an indication that the signal is actually caused by an eclipsing binary at twice the estimated orbital period. Although {\tt vespa} accounts for this scenario in its FPP calculation, we do not validate the candidate even though its FPP is below 1\%. The candidate 201180665.01 has a relatively high FPP ($\sim$64\%), and also a suspiciously large radius estimate ($\sim$26 \rearth). Although spectroscopic characterization could yield a different radius estimate for the host star (and thus also for the candidate), we conclude that this is most likely an eclipsing M dwarf companion. The candidates 228974907.01, and 228846243.01 do not have particularly low FPPs, but they may be interesting targets for further observations due to their relatively long orbital periods. The candidate 201128338.01 was statistically validated previously in the literature \citep[K2-152\,b;][]{2018AJ....155..127H}; we find a similarly low FPP, but we do not validate it simply because it has fewer than three transits in the \ktwo photometry (and thus odd/even variations in transit depth cannot be robustly ruled out). Further observations will shed light on the true nature of these candidates, either by measuring RV variations with precision spectrographs or via simultaneous multi-band transit observations with instruments such as MuSCAT \citep{2015JATIS...1d5001N} and MuSCAT2 (a {\it griz} clone of MuSCAT now in operation at Teide Observatory).

The integrated FPP is $\sim$2.1 for the full set of \numtotal candidates, which implies the existence of two false positives in the sample. We have already confirmed that 228729473.01 is a false positive via RV observations (see Csizmadia et al., in prep.), and we suspect 229133720.01, 201390048.01, and 201180665.01 of being false positives, as described above. Therefore, we expect no false positives among the remainder of the sample, and most of the \numcand unvalidated candidates could be statistically validated or confirmed by future observations.

\subsection{Interesting new systems}

\subsubsection{Ultra-short period planets}

Ultra-short period planets (USPs) are defined by having orbital periods less than one day \citep[e.g.][]{2013ApJ...774...54S, 2015ApJ...812..112S}. Our validated planet sample contains \nusp USPs: K2-131\,b \citep{2017AJ....154..226D}; K2-156\,b and K2-223\,b \citep{2018AJ....155..136M}; and K2-229\,b \citep{2018AJ....155..136M, 2018NatAs...2..393S}. These planets join a growing list of USPs discovered by \ktwo \citep[e.g.][]{2016ApJ...829L...9V, 2017AJ....154..122C, 2017AJ....154..123G, 2017AJ....153...82A, 2017AJ....154..226D, 2017arXiv171102097B, 2018arXiv180103502M}. The radii of these USPs place all three of them below the recently observed gap in the radius distribution \citep{2017AJ....154..109F, 2017arXiv171005398V} which was predicted as a consequence of photoevaporation \citep[e.g.][]{2013ApJ...775..105O, 2014ApJ...792....1L}. These three USPs are therefore likely to be rocky and have high densities, consistent with having lost any primordial or secondary atmospheres they might once have had. Of these validated USPs, we measured the metallicity of the host stars spectroscopically for three of them; K2-229 appears to have only a modestly sub-solar metallicity of $-0.09 \pm 0.02$ \feh, but K2-131 and K2-156 have more significantly sub-solar metallicities of $-0.17 \pm 0.03$ and $-0.25 \pm 0.06$ \feh, respectively (see \autoref{tab:stellar}). Due to their small size, these USPs are likely to have a mass less than 5--6 \mearth, and thus the sub-solar metallicity of their host stars would be consistent with the USP mass-metallicity trend noted by \citet{2017AJ....153..271S} (i.e. similar to Kepler-78\,b and Kepler-10\,b).

The G dwarf K2-223 and K dwarf K2-229 are both relatively bright ($Kp \sim 11$ mag), and host planets with predicted masses and Doppler semi-amplitudes well within the reach of current precision spectrographs, such as HARPS or HIRES. K2-156\,b orbits a slightly fainter star and has a slightly smaller predicted mass and Doppler semi-amplitude, but is also a viable target for characterization with today's instrumentation. Such mass measurements would yield densities and constrain the bulk compositions of these USPs, which would enable tests of USP formation theories.

In addition to the \nusp validated USPs mentioned above, we also note that our sample contains two USP candidates: 201595106.01 and 228836835.01. We do not validate 201595106.01 because of the presence of a faint star in the EPIC with a $\Delta Kp$ of 5.839 and a separation of 13.62\arcsec (see \autoref{tab:epic-companions}), which is within the photometric aperture we used to extract the \ktwo light curve. We do not validate 228836835.01 because it has a FPP of $\sim$4\% and thus does not meet our validation criterion. Future observations could potentially rule out false positive scenarios for both of these candidates, resulting in the validation of two more USPs from \ktwo C10.

\subsubsection{Multi-planet systems}

Of the \numvalid validated planets in our sample, \nummulti of them were found in two-planet systems, which enables the study of their orbital architectures and evolution. Four of these systems have orbital architectures with period ratios just wide of a 2:1 commensurability, and two are close to a 3:1 commensurability. The pairs closest to 2:1 are K2-243\,bc and K2-154\,bc, which both have $P_c / P_b \approx 2.16$. The relatively large fraction of multi-planet systems (4/9) in our sample with period ratios just wide of a 2:1 commensurability is reminiscent of the distribution of orbital architectures observed with \kepler \citep{2014ApJ...790..146F}. K2-254\,bc and K2-247\,bc are both just inside a 3:1 commensurability, with period ratios of $P_c / P_b \approx 2.96$ and $P_c / P_b \approx 2.89$, respectively. Although we did not detect any significant TTVs in the \ktwo data, some of these systems may have TTVs which could be detected with higher cadence transit observations.

Intriguingly, two of the \nusp validated USPs in the sample were found in two-planet systems with large period ratios, similar to the Kepler-10 system: K2-223\,bc has $P_c / P_b \approx 9.02$, and K2-229\,bc has $P_c / P_b \approx 14.25$. The presence of an additional transiting planet decreases the likelihood that these USPs reached their current orbits via dynamical scattering, as this would increase the chances of higher mutual inclinations; even after tidal circularization, the geometric transit probability would be decreased by a higher likelihood of non-coplanarity. This is consistent with previous analyses in which USP systems have been noted to be dynamically cold \citep[e.g.][]{2017AJ....154..226D}.

\subsection{Characterization targets}
\label{sec:characterization}

We predicted the masses of the candidates using the probabilistic mass-radius relation of \citet{2016ApJ...825...19W}\footnote{\url{https://github.com/dawolfgang/MRrelation}} (see \autoref{tab:params}). The predicted masses enabled us to compute other quantities of interest, which we then used to identify potentially interesting targets for follow-up characterization via Doppler and transmission spectroscopy.

\subsubsection{Doppler targets}

We computed the expected Doppler semi-amplitude due to the reflex motion of the host star induced by each planet (see \autoref{tab:params}). We used these expected semi-amplitudes in conjunction with the brightness of the host stars to identify planets in the sample which are good targets for radial velocity (RV) follow-up study using current and future facilities. Such RV observations will reveal the planets' densities and constrain their bulk compositions. This is of particular interest for relatively small planets with radii in the range $1.5 - 2.5$ \rearth because such measurements could enable tests of planet formation theories and post-processes, such as the photoevaporation \citep[e.g.][]{2013ApJ...775..105O, 2014ApJ...792....1L}, which has been proposed to explain the observed gap in the radius distribution \citep{2017AJ....154..109F,2017arXiv171005398V}. However, because of the difficulty of detecting the small Doppler signals of such planets, it is especially important to identify such planets which are orbiting relatively bright stars, for which the RV precision required to measure their masses is more readily obtainable. \autoref{tab:rv} lists validated planets with predicted Doppler semi-amplitudes greater than 1 m\,s$^{-1}$ orbiting stars brighter than $Kp$ = 12 mag. For convenience, we also list planetary orbital periods and stellar rotational periods (when available); potentially confounding quasi-periodic RV signals produced by stellar magnetic activity are less likely to present a challenge for mass measurement when the orbital period is far from the stellar rotational period (or a harmonic). We note that 228732031.01 (K2-131\,b) and 228801451.01 (K2-229\,b) both already have measured masses via precision RVs \citep{2017AJ....154..226D, 2018NatAs...2..393S}.

Another possibly interesting RV target is K2-257\,b, a sub-Earth-size planet orbiting a nearby M dwarf. Although the planet's radius is only $0.83^{+0.06}_{-0.05}$ \rearth, the Doppler semi-amplitude could be as high as $\sim$1 m\,s$^{-1}$ due to the low mass of the host star and the planet's short orbital period. The host star is moderately bright ($Kp = 12.873$, $J = 10.477$ mag), so this presents an opportunity to directly measure the mass of a sub-Earth with one of today's high precision optical or NIR spectrographs. Such a measurement would yield the planet's density and constrain its composition, as well as improve our knowledge of the mass-radius relation for small planets. The only other sub-Earth-size planet known to transit a similarly bright M dwarf is Kepler-138\,b, for which the mass has been measured only via transit timing variations \citep{2015Natur.522..321J,2018MNRAS.478..460A}.

\begin{deluxetable}{lllllc}
\tabletypesize{\scriptsize}
\tablecaption{Validated planets with predicted Doppler semi-amplitudes greater than 1 m\,s$^{-1}$ orbiting stars brighter than $Kp$ = 12 mag. Note: 228721452.01 is not listed here because it doesn't meet these criteria, but RV measurements to constrain the mass of 228721452.02 could also reveal the inner planet's mass, as both Keplerian signals would need to be accounted for in the RV analysis. \label{tab:rv}}
\tablehead{EPIC & $Kp$ & $K_\mathrm{pred}$ & \rp & $P_\mathrm{orb}$ & $P_\mathrm{rot}$ \\
       & [mag]& [m\,s$^{-1}$] & [\rearth] & [days] & [days] }
\startdata
\input{tab_rv.tex}
\enddata
\end{deluxetable}

\subsubsection{Atmospheric targets}

In order to identify viable new targets for atmospheric studies via transmission spectroscopy, we used the properties of the host stars and planets to predict atmospheric scale heights and the amplitudes of the wavelength dependence of transit depth ($\delta_{\mathrm{TS}}$).
Following \citet{2009ApJ...690.1056M}, we calculated the atmospheric scale height $H$ and $\delta_{\mathrm{TS}}$ for each validated planet by
\begin{eqnarray}
H &=& \frac{29.26}{(\mu/28.96)}\frac{T_\mathrm{eq}}{g} ~~~[\mathrm{m}]\\
\delta_{\mathrm{TS}} &\sim & 10\,H \cdot \rp/\rstar^2,
\end{eqnarray}
where $\mu$, $T_\mathrm{eq}$, and $g$ are the mean molecular weight, planet equilibrium
temperature, and planet surface gravity, respectively.
We used the predicted planet mass estimated in \autoref{sec:characterization} to predict the surface gravity, and assumed a bond albedo of 0.3 and a mean molecular weight $\mu = 2$ (hydrogen-dominated atmosphere) for each planet (see \autoref{tab:atmospheric}). We note that this assumption for $\mu$ is likely to be invalid for the smaller planets in our sample (i.e. \rp $\lesssim$ 1.5--2 \rearth), as they are not likely to have substantial hydrogen-dominated atmospheres; these smaller planets likely have higher mean molecular weight atmospheres, which would make their characterization via transmission spectroscopy more challenging. The validated planets K2-140\,b and K2-255\,b both orbit relatively bright host stars ($J < 12$ mag) and have large expected transmission spectroscopy signals ($\delta_{\mathrm{TS}} > 200$ ppm), and thus could be interesting targets for future atmospheric characterization.

\section{Summary}
\label{sec:summary}

We detected \numtotal planet candidates in \ktwo Campaign 10 and obtained high resolution imaging and spectroscopy follow-up observations to characterize the host stars. We performed detailed modeling of the light curves and used the resulting transit parameters to compute physical planet properties. We used the planet and host star properties to predict masses and atmospheric signals, which enabled us to identify good targets for future characterization via Doppler and transmission spectroscopy. We statistically validated \numvalid planets, leaving a remainder of \numcand candidates and one false positive. We expect nearly all of these remaining candidates to be real planets, which could potentially be validated via further observations and analysis.

\acknowledgements
This work was carried out as part of the KESPRINT consortium. The WIYN/NESSI observations were conducted as part of an approved NOAO observing program (P.I. Livingston, proposal ID 2017A-0377). Data presented herein were obtained at the WIYN Observatory from telescope time allocated to NN-EXPLORE through the scientific partnership of the National Aeronautics and Space Administration, the National Science Foundation, and the National Optical Astronomy Observatory. This work was supported by a NASA WIYN PI Data Award, administered by the NASA Exoplanet Science Institute. NESSI was funded by the NASA Exoplanet Exploration Program and the NASA Ames Research Center. NESSI was built at the Ames Research Center by Steve B. Howell, Nic Scott, Elliott P. Horch, and Emmett Quigley. The authors are honored to be permitted to conduct observations on Iolkam Du'ag (Kitt Peak), a mountain within the Tohono O'odham Nation with particular significance to the Tohono O'odham people. J.\,H.\,L. gratefully acknowledges the support of the Japan Society for the Promotion of Science (JSPS) Research Fellowship for Young Scientists. This work was supported by Japan Society for Promotion of Science (JSPS) KAKENHI Grant Number JP16K17660. M.E. and W.D.C. were supported by NASA grant NNX16AJ11G to The University of Texas. This paper includes data collected by the \kepler\ mission. Funding for the \kepler\ mission is provided by the NASA Science Mission directorate.

\facilities{Kepler, WIYN (NESSI), McDonald (Tull), NOT (FIES), TNG (HARPS-N)}

\software{{\tt scipy}, {\tt emcee}, {\tt batman}, {\tt vespa}, {\tt IRAF}, {\tt pyaneti}, {\tt exotrending}}

\bibliography{ref.bib}

\begin{longrotatetable}
\begin{deluxetable*}{llllllllllllllc}
\tabletypesize{\tiny}
\small
\setlength\tabcolsep{3pt}
\tablecaption{Planet and candidate parameters. \label{tab:params}}
\tablehead{EPIC & Name & $P_\mathrm{orb}$ & $T_0$ & $a$ & $b$ & \rp & $a$ & \rp & \teq & $M_P$ & $K_\mathrm{pred}$ & $\rho_{\star,LC}$ & FPP & Disposition \\
  	& & [days] & [BKJD] & [\rstar] &  & [\rstar] & [AU] & [\rearth] & [K] & [\mearth] & [m\,s$^{-1}$] & [$\rho_\odot$] & & }
\startdata
\input{tab_params.tex}
\enddata
\tablecomments{$M_P$ is the mass predicted using the mass-radius relation of \citet{2016ApJ...825...19W} (see \autoref{sec:characterization}). We note that this mass-radius relation was calibrated with sub-Neptunes similar in size to the vast majority of the planets in our validated sample; the predictions may not be accurate for larger candidates, but we report them here anyway for the sake of uniformity. The ``Disposition" column indicates the final validation status of each candidate: ``VP'' = validated planet; ``PC'' = planet candidate; ``FP'' = false positive.}
\end{deluxetable*}
\end{longrotatetable}

\startlongtable
\begin{deluxetable*}{llllllllr}
\tablecaption{Stellar parameters. \label{tab:stellar}}
\tablehead{EPIC & $T_{\mathrm{eff}}$ & log $g$ & [Fe/H] & Mass & Radius & Distance & \vsini & Provenance\\
       & [K] & [cgs] & [dex] & [$M_{\odot}$] & [$R_{\odot}$] & [pc] & [km/s] & }
\startdata
\input{tab_stellar.tex}
\enddata
\tablecomments{``Provenance'' indicates the source of the spectroscopic parameters used as priors in our analysis (see \autoref{sec:stellar}). The \vsini uncertainties are internal to the {\it Kea} pipeline and do not account for other types of line broadening; thus they are likely to be underestimated.}
\end{deluxetable*}

\startlongtable
\begin{deluxetable*}{lllllllll}
\tablecaption{Individual false positive scenario likelihoods computed by {\tt vespa}. Note: $a$ -- likelihood that the signal is due to a background eclipsing binary, at the measured period or twice that; $b$ -- likelihood that the signal is due to an eclipsing binary, at the measured period or twice that; $c$ -- likelihood that the signal is due to a hierarchical star system with an eclipsing component, at the measured period or twice that; $d$ -- likelihood that the signal is due to a planet. \label{tab:vespa}}
\tablehead{EPIC &     L\_beb$^a$ &  L\_beb\_Px2$^a$ &          L\_eb$^b$ &      L\_eb\_Px2$^b$ &         L\_heb$^c$ &     L\_heb\_Px2$^c$ &     L\_pl$^d$ &     FPP}
\startdata
\input{tab_vespa.tex}
\enddata
\end{deluxetable*}

\startlongtable
\begin{deluxetable}{lcc}
\tablecaption{Additional constraints to {\tt vespa}. The columns ``maxrad'' and ``secthresh'' refer to the maximum radius (the angular size of the photometric aperture) and the secondary eclipse threshold (the maximum secondary eclipse depth allowed by the light curve), respectively. \label{tab:vespa_constraints}}
\tablehead{EPIC & maxrad [arcsec] & secthresh}
\startdata
\input{tab_vespa_constraints.tex}
\enddata
\end{deluxetable}

\startlongtable
\begin{deluxetable}{lccr}
\tablecaption{WIYN/NESSI datasets used in this work. \label{tab:speckle_obs}}
\tablehead{ EPIC & Filter center & Filter width &    Obs. Date \\
    & [nm] & [nm] & }
\startdata
\input{tab_speckle_obs.tex}
\enddata
\end{deluxetable}

\begin{deluxetable*}{lllllllll}
\tablecaption{TNG/HARPS-N results. \label{tab:harps}}
\tablehead{EPIC & $T_\mathrm{obs} $ & RV & BIS & FWHM & log(RHK) & B-V & $T_\mathrm{exp}$ & SNR \\
 & [BJD$_\mathrm{TDB}$] & [km\,s$^{-1}$] & [km\,s$^{-1}$] & [km\,s$^{-1}$] & & [mag] & [sec] & [5500 nm] }
\startdata
\input{tab_harps.tex}
\enddata
\end{deluxetable*}

\startlongtable
\begin{deluxetable}{llll}
\tablecaption{Predicted atmospheric characteristics, where g is surface gravity, H is atmospheric scale height, and $\delta_{\mathrm{TS}}$ is the expected amplitude of atmospheric spectral features. \label{tab:atmospheric}}
\tablehead{EPIC & $g$ & H & $\delta_{\mathrm{TS}}$ \\
       & [$g_\oplus$] & [km] & [ppm]}
\startdata
\input{tab_atmospheric.tex}
\enddata
\end{deluxetable}

\startlongtable
\begin{deluxetable*}{lllllllll}
\tablecaption{Comparison of parameters between \ktwo pipelines. \label{tab:comparison}}
\tablehead{EPIC & $P_\mathrm{orb}$ & $\Delta P$ & \rp & $\Delta$\rp & $b$ & $\Delta b$ & $a$ & $\Delta a$ \\
  & [days] & [$\sigma$] & [\rstar] & [$\sigma$] & & [$\sigma$] & [\rstar] & [$\sigma$] }
\startdata
\input{tab_comparison.tex}
\enddata
\end{deluxetable*}

\end{document}

%% file: tab_candidates.tex
201092629 &   11.9 & 26.810 & 2751.22 &   4.1 &  0.00090 & 13.2 &     22$^{+6}_{-2}$ \\
201102594 &   15.6 &  6.514 & 2753.24 &   2.0 &  0.00624 &  8.2 &      25$\pm$3 \\
201110617 &   12.9 &  0.813 & 2750.14 &   1.3 &  0.00029 & 16.2 &  16.8$\pm$2.5 \\
201111557 &   11.4 &  2.302 & 2750.17 &   1.9 &  0.02268 &  7.6 &  12.0$\pm$1.8 \\
201127519 &   11.6 &  6.179 & 2752.55 &   2.5 &  0.01303 & 11.6 &        --- \\
201128338 &   13.1 & 32.655 & 2775.62 &   4.0 &  0.00159 &  6.7 &  15.6$\pm$2.2 \\
201132684 &   11.7 & 10.061 & 2757.49 &   3.8 &  0.00070 &  8.7 &  13.8$\pm$1.3 \\
201132684 &   11.7 &  5.906 & 2750.82 &   5.0 &  0.00015 &  9.7 &  13.8$\pm$1.3 \\
201164625 &   11.9 &  2.711 & 2750.15 &   3.1 &  0.00020 &  6.7 &  12.5$\pm$1.5 \\
201166680 &   10.9 & 24.941 & 2751.51 &   5.2 &  0.00019 &  6.6 &        --- \\
201166680 &   10.9 & 11.540 & 2760.22 &   3.7 &  0.00016 &  7.8 &        --- \\
201180665 &   13.1 & 17.773 & 2753.50 &   2.9 &  0.03662 & 11.2 &        --- \\
201211526 &   11.7 & 21.070 & 2755.48 &   3.9 &  0.00030 &  8.3 &        --- \\
201225286 &   11.7 & 12.420 & 2753.52 &   3.3 &  0.00065 & 11.6 &  20.8$\pm$1.6 \\
201274010 &   13.9 & 13.008 & 2756.51 &   2.2 &  0.00065 &  7.7 &        --- \\
201352100 &   12.8 & 13.383 & 2761.79 &   2.2 &  0.00120 & 12.5 &     36$\pm$11 \\
201357643 &   12.0 & 11.893 & 2754.55 &   4.2 &  0.00107 & 12.3 &        --- \\
201386739 &   14.4 &  5.767 & 2750.70 &   3.4 &  0.00134 & 11.1 &      35$\pm$6 \\
201390048 &   12.0 &  9.455 & 2750.92 &   3.0 &  0.02669 &  7.7 &        --- \\
201390927 &   14.2 &  2.638 & 2750.34 &   1.7 &  0.00110 & 12.9 &        --- \\
201392505 &   13.4 & 27.463 & 2759.08 &   5.5 &  0.00150 &  9.3 &        --- \\
201437844 &    9.2 & 21.057 & 2757.07 &   4.4 &  0.00100 & 10.0 &        --- \\
201437844 &    9.2 &  9.560 & 2753.52 &   3.5 &  0.00030 &  9.8 &        --- \\
201595106 &   11.7 &  0.877 & 2750.05 &   1.0 &  0.00025 &  9.4 &        --- \\
201598502 &   14.3 &  7.515 & 2755.43 &   2.3 &  0.00129 &  7.5 &        --- \\
201615463 &   12.0 &  8.527 & 2753.77 &   3.7 &  0.00016 &  7.2 &        --- \\
228707509 &   14.8 & 15.351 & 2752.51 &   3.6 &  0.02386 & 13.6 &        --- \\
228720681 &   13.8 & 15.782 & 2753.42 &   3.4 &  0.01028 & 14.3 &   9.8$\pm$1.1 \\
228721452 &   11.3 &  4.563 & 2749.98 &   2.8 &  0.00020 & 12.6 &        --- \\
228721452 &   11.3 &  0.506 & 2750.56 &   0.9 &  0.00010 &  9.6 &        --- \\
228724899 &   13.3 &  5.203 & 2753.45 &   1.4 &  0.00113 & 12.3 &        --- \\
228725791 &   14.3 &  6.492 & 2755.15 &   1.7 &  0.00110 &  9.8 &      32$\pm$3 \\
228725791 &   14.3 &  2.251 & 2749.97 &   1.2 &  0.00100 &  7.3 &      32$\pm$3 \\
228725972 &   12.5 &  4.477 & 2752.69 &   2.4 &  0.03270 & 11.5 &        --- \\
228725972 &   12.5 & 10.096 & 2755.41 &   3.6 &  0.05928 & 13.0 &        --- \\
228729473 &   11.5 & 16.773 & 2752.76 &  12.4 &  0.00199 & 11.6 &     36$^{+5}_{-3}$ \\
228732031 &   11.9 &  0.369 & 2749.93 &   1.0 &  0.00040 & 15.1 &   9.4$\pm$1.9 \\
228734900 &   11.5 & 15.872 & 2754.37 &   4.6 &  0.00034 &  8.0 &        --- \\
228735255 &   12.5 &  6.569 & 2755.29 &   3.3 &  0.01280 & 12.6 &  31.1$\pm$2.0 \\
228736155 &   12.0 &  3.271 & 2751.02 &   2.4 &  0.00027 &  9.3 &        --- \\
228739306 &   13.3 &  7.172 & 2755.11 &   2.8 &  0.00070 &  8.1 &        --- \\
228748383 &   12.5 & 12.409 & 2750.04 &   5.9 &  0.00024 &  8.0 &        --- \\
228748826 &   13.9 &  4.014 & 2751.13 &   2.4 &  0.00102 & 13.2 &     39$^{+6}_{-8}$ \\
228753871 &   13.2 & 18.693 & 2757.74 &   2.2 &  0.00082 &  7.7 &  16.4$\pm$2.3 \\
228758778 &   14.8 &  9.301 & 2756.07 &   2.7 &  0.00214 &  7.8 &        --- \\
228758948 &   12.9 & 12.203 & 2753.83 &   4.0 &  0.00128 & 12.4 &  11.3$\pm$1.7 \\
228763938 &   12.6 & 13.814 & 2763.19 &   3.6 &  0.00036 &  8.8 &        --- \\
228784812 &   12.6 &  4.189 & 2751.02 &   2.2 &  0.00014 &  8.9 &        --- \\
228798746 &   12.7 &  2.697 & 2750.20 &   1.5 &  0.02587 & 14.1 &        --- \\
228801451 &   11.0 &  8.325 & 2753.35 &   2.5 &  0.05325 & 12.9 &  19.5$\pm$2.7 \\
228801451 &   11.0 &  0.584 & 2750.46 &   1.5 &  0.01625 & 10.0 &  19.5$\pm$2.7 \\
228804845 &   12.6 &  2.860 & 2749.60 &   2.6 &  0.00020 &  7.3 &        --- \\
228809391 &   12.6 & 19.580 & 2763.80 &   2.6 &  0.00100 &  8.3 &        --- \\
228809550 &   14.7 &  4.002 & 2751.00 &   2.1 &  0.01259 & 12.5 &        --- \\
228834632 &   14.9 & 11.730 & 2758.63 &   2.1 &  0.00111 &  8.6 &  23.6$\pm$2.1 \\
228836835 &   14.9 &  0.728 & 2750.26 &   0.8 &  0.00068 & 15.4 &        --- \\
228846243 &   14.5 & 25.554 & 2756.93 &   5.4 &  0.00220 & 10.5 &        --- \\
228849382 &   13.8 & 12.120 & 2757.61 &   2.4 &  0.00120 &  7.6 &        --- \\
228849382 &   13.8 &  4.097 & 2749.96 &   1.6 &  0.00052 &  8.8 &        --- \\
228888935 &   14.1 &  5.691 & 2751.67 &   3.3 &  0.00533 & 10.3 &   7.2$\pm$1.1 \\
228894622 &   13.3 &  1.964 & 2750.31 &   1.1 &  0.00183 & 16.3 &  20.8$\pm$2.4 \\
228934525 &   13.4 &  3.676 & 2752.05 &   1.7 &  0.00110 & 14.2 &  28.3$\pm$3.1 \\
228934525 &   13.4 &  7.955 & 2751.34 &   2.1 &  0.00110 & 11.4 &  28.3$\pm$3.1 \\
228964773 &   14.9 & 37.209 & 2776.76 &   3.1 &  0.00280 &  6.9 &        --- \\
228968232 &   14.7 &  5.520 & 2753.52 &   3.6 &  0.00097 &  8.6 &        --- \\
228974324 &   12.9 &  1.606 & 2750.29 &   1.3 &  0.00034 & 13.1 &  22.0$\pm$2.3 \\
228974907 &    9.3 & 20.782 & 2759.64 &   5.0 &  0.00010 &  7.2 &        --- \\
229004835 &   10.2 & 16.138 & 2764.63 &   2.1 &  0.00036 & 10.6 &  22.2$\pm$2.5 \\
229017395 &   13.2 & 19.099 & 2753.28 &   6.0 &  0.00049 &  8.1 &        --- \\
229103251 &   13.7 & 11.667 & 2756.72 &   3.1 &  0.00114 &  9.9 &        --- \\
229131722 &   12.5 & 15.480 & 2752.71 &   4.2 &  0.00037 &  8.3 &        --- \\
229133720 &   11.5 &  4.037 & 2750.96 &   1.5 &  0.00091 & 12.4 &  11.8$\pm$1.3 \\

%% file: tab_epic_companions.tex
201111557 &         201111694 &              15.90 &                5.187 \\
201164625 &         201164669 &              17.58 &                3.228 \\
201595106 &         201595004 &              13.62 &                5.839 \\
228707509 &         228707572 &              12.48 &                1.563 \\
228720681 &         228720649 &               7.86 &                2.905 \\
228758948 &         228758983 &               9.00 &                3.267 \\

%% file: tab_rv.tex
201092629.01  &  11.858  &  $2.5^{+0.7}_{-0.7}$  &  2.55  &  26.8199  &     22$^{+6}_{-2}$  \\
201132684.01  &  11.678  &  $1.3^{+0.7}_{-0.7}$  &  1.28  &   5.9028  &  13.8$\pm$1.3  \\
201132684.02  &  11.678  &  $3.0^{+0.8}_{-0.8}$  &  2.64  &  10.0605  &  13.8$\pm$1.3  \\
201166680.01  &  10.897  &  $1.8^{+0.6}_{-0.6}$  &  2.17  &  11.5418  &        ---  \\
201166680.02  &  10.897  &  $1.2^{+0.5}_{-0.4}$  &  2.01  &  24.9460  &        ---  \\
201211526.01  &  11.696  &  $1.5^{+0.6}_{-0.6}$  &  1.75  &  21.0688  &        ---  \\
201225286.01  &  11.729  &  $2.3^{+0.7}_{-0.7}$  &  2.26  &  12.4220  &  20.8$\pm$1.6  \\
201357643.01  &  11.998  &  $5.7^{+1.2}_{-1.1}$  &  4.34  &  11.8931  &        ---  \\
201437844.01  &   9.234  &  $2.3^{+0.7}_{-0.7}$  &  2.32  &   9.5580  &        ---  \\
201437844.02  &   9.234  &  $3.9^{+0.8}_{-0.7}$  &  4.31  &  21.0579  &        ---  \\
201615463.01  &  11.964  &  $2.1^{+0.7}_{-0.7}$  &  2.19  &   8.5270  &        ---  \\
228721452.02  &  11.325  &  $1.8^{+0.9}_{-0.8}$  &  1.57  &   4.5633  &        ---  \\
228732031.01  &  11.937  &  $5.3^{+2.3}_{-2.2}$  &  1.70  &   0.3693  &   9.4$\pm$1.9  \\
228734900.01  &  11.535  &  $2.9^{+0.7}_{-0.6}$  &  3.49  &  15.8721  &        ---  \\
228801451.01  &  10.955  &  $2.2^{+1.0}_{-1.2}$  &  1.14  &   0.5843  &  19.5$\pm$2.7  \\
228801451.02  &  10.955  &  $2.3^{+0.8}_{-0.8}$  &  2.03  &   8.3273  &  19.5$\pm$2.7  \\

%% file: tab_params.tex
201092629.01     &      K2-241\,b     &  $26.81990^{+0.00245}_{-0.00247}$     &  $2751.2063^{+0.0035}_{-0.0034}$     &    $43.3^{+3.8}_{-9.9}$     &  $0.41^{+0.30}_{-0.28}$     &  $0.0312^{+0.0023}_{-0.0012}$     &  $0.1567^{+0.0007}_{-0.0008}$     &   $2.55^{+0.18}_{-0.10}$     &       $507^{+5}_{-5}$     &       $9.1^{+2.4}_{-2.4}$     &     $2.4^{+0.7}_{-0.7}$     &   $1.5^{+0.4}_{-0.8}$     &   $7 \times 10^{-8}$     &          VP  \\
201102594.01     &      K2-242\,b     &   $6.51389^{+0.00082}_{-0.00083}$     &  $2753.2400^{+0.0054}_{-0.0054}$     &    $22.1^{+3.8}_{-5.6}$     &  $0.42^{+0.32}_{-0.29}$     &  $0.0633^{+0.0051}_{-0.0044}$     &  $0.0494^{+0.0004}_{-0.0004}$     &   $2.54^{+0.21}_{-0.19}$     &       $416^{+8}_{-8}$     &       $9.0^{+2.7}_{-2.5}$     &     $5.9^{+1.7}_{-1.6}$     &   $3.4^{+2.1}_{-2.0}$     &  $6 \times 10^{-12}$     &          VP  \\
201110617.01     &      K2-156\,b     &   $0.81314^{+0.00007}_{-0.00008}$     &  $2750.1427^{+0.0041}_{-0.0038}$     &     $4.1^{+0.6}_{-0.9}$     &  $0.41^{+0.32}_{-0.27}$     &  $0.0161^{+0.0014}_{-0.0010}$     &  $0.0149^{+0.0001}_{-0.0001}$     &   $1.15^{+0.10}_{-0.07}$     &    $1347^{+18}_{-19}$     &       $2.7^{+1.5}_{-1.5}$     &     $2.4^{+1.3}_{-1.4}$     &   $1.4^{+0.7}_{-0.7}$     &  $2 \times 10^{-12}$     &          VP  \\
201111557.01     &                    &   $2.30183^{+0.00028}_{-0.00030}$     &  $2750.1688^{+0.0052}_{-0.0052}$     &    $11.8^{+2.1}_{-3.0}$     &  $0.42^{+0.33}_{-0.28}$     &  $0.0144^{+0.0014}_{-0.0010}$     &  $0.0313^{+0.0003}_{-0.0004}$     &   $1.12^{+0.11}_{-0.08}$     &    $1054^{+55}_{-55}$     &       $2.4^{+1.6}_{-1.4}$     &     $1.4^{+0.9}_{-0.8}$     &   $4.1^{+2.6}_{-2.4}$     &   $3 \times 10^{-4}$     &          PC  \\

%% file: tab_stellar.tex
201092629  &    $5262^{+43}_{-39}$  &         $4.54 \pm 0.01$  &  $-0.44^{+0.04}_{-0.03}$  &         $0.71 \pm 0.01$  &         $0.75 \pm 0.01$  &     $149.6^{+1.1}_{-1.0}$  &   $2.08 \pm 0.29$  &              This work  \\
201102594  &    $3459^{+65}_{-38}$  &         $4.89 \pm 0.01$  &  $-0.01^{+0.16}_{-0.18}$  &         $0.38 \pm 0.01$  &         $0.37 \pm 0.01$  &           $109.6 \pm 0.6$  &               ---  &                    ---  \\
201110617  &         $4597 \pm 50$  &  $4.62^{+0.02}_{-0.01}$  &         $-0.25 \pm 0.06$  &  $0.66^{+0.02}_{-0.01}$  &         $0.66 \pm 0.01$  &           $150.4 \pm 1.0$  &   $1.80 \pm 0.30$  &              This work  \\
201111557  &  $5011^{+289}_{-239}$  &         $4.62 \pm 0.01$  &         $-0.23 \pm 0.23$  &  $0.77^{+0.02}_{-0.03}$  &         $0.71 \pm 0.01$  &            $97.5 \pm 0.5$  &               ---  &                    ---  \\
201127519  &    $4957^{+47}_{-49}$  &  $4.58^{+0.02}_{-0.03}$  &          $0.03 \pm 0.05$  &         $0.81 \pm 0.03$  &         $0.77 \pm 0.01$  &           $118.0 \pm 0.7$  &   $1.85 \pm 0.22$  &              This work  \\
201128338  &    $4044^{+34}_{-35}$  &         $4.67 \pm 0.01$  &   $0.12^{+0.09}_{-0.08}$  &         $0.63 \pm 0.01$  &         $0.61 \pm 0.01$  &           $108.7 \pm 0.4$  &               ---  &     Hirano et al. 2018  \\
201132684  &    $5503^{+51}_{-48}$  &         $4.45 \pm 0.02$  &          $0.07 \pm 0.07$  &  $0.92^{+0.03}_{-0.02}$  &         $0.95 \pm 0.02$  &           $198.2 \pm 2.1$  &               ---  &       Mayo et al. 2018  \\
201164625  &    $6264^{+83}_{-81}$  &         $3.66 \pm 0.03$  &   $0.18^{+0.02}_{-0.03}$  &  $1.78^{+0.06}_{-0.03}$  &  $3.30^{+0.13}_{-0.16}$  &  $1067.1^{+36.4}_{-53.7}$  &  $11.50 \pm 0.46$  &              This work  \\

%% file: tab_vespa.tex
201092629.01     &                   $0$     &                   $0$     &   $1.1 \times 10^{-11}$     &  $7.8 \times 10^{-10}$     &   $5.7 \times 10^{-52}$     &   $1.4 \times 10^{-20}$     &  $1.2 \times 10^{-2}$     &   $6.8 \times 10^{-8}$  \\
201102594.01     &                   $0$     &                   $0$     &   $1.0 \times 10^{-22}$     &  $1.1 \times 10^{-13}$     &   $7.6 \times 10^{-30}$     &   $2.9 \times 10^{-21}$     &  $1.8 \times 10^{-2}$     &  $5.9 \times 10^{-12}$  \\
201110617.01     &                   $0$     &                   $0$     &   $2.3 \times 10^{-52}$     &  $4.5 \times 10^{-13}$     &   $1.1 \times 10^{-90}$     &   $3.7 \times 10^{-42}$     &  $2.3 \times 10^{-1}$     &  $2.0 \times 10^{-12}$  \\
201111557.01     &                   $0$     &                   $0$     &    $6.1 \times 10^{-9}$     &   $1.2 \times 10^{-6}$     &   $1.0 \times 10^{-32}$     &   $2.4 \times 10^{-14}$     &  $4.0 \times 10^{-3}$     &   $3.0 \times 10^{-4}$  \\
201127519.01     &                   $0$     &                   $0$     &    $1.1 \times 10^{-3}$     &   $3.2 \times 10^{-6}$     &   $2.2 \times 10^{-86}$     &   $4.6 \times 10^{-27}$     &  $2.8 \times 10^{-2}$     &   $3.9 \times 10^{-2}$  \\
201128338.01     &                   $0$     &                   $0$     &   $3.0 \times 10^{-16}$     &  $2.0 \times 10^{-10}$     &   $3.0 \times 10^{-12}$     &   $1.1 \times 10^{-12}$     &  $1.6 \times 10^{-3}$     &   $1.3 \times 10^{-7}$  \\
201132684.01     &                   $0$     &                   $0$     &   $4.7 \times 10^{-14}$     &   $3.4 \times 10^{-9}$     &   $4.4 \times 10^{-25}$     &   $2.3 \times 10^{-12}$     &  $4.7 \times 10^{-3}$     &   $7.2 \times 10^{-7}$  \\
201132684.02     &                   $0$     &                   $0$     &    $2.1 \times 10^{-9}$     &   $5.6 \times 10^{-8}$     &   $1.2 \times 10^{-49}$     &   $1.4 \times 10^{-18}$     &  $2.9 \times 10^{-2}$     &   $2.0 \times 10^{-6}$  \\

%% file: tab_vespa_constraints.tex
201092629.01     &         20.7     &  1$ \times 10^{-4}$  \\
201102594.01     &         15.1     &  2$ \times 10^{-4}$  \\
201110617.01     &         15.9     &  1$ \times 10^{-4}$  \\
201111557.01     &         17.5     &  1$ \times 10^{-4}$  \\
201127519.01     &         10.3     &  3$ \times 10^{-3}$  \\
201128338.01     &         18.3     &  2$ \times 10^{-4}$  \\
201132684.01     &         21.5     &  2$ \times 10^{-4}$  \\
201132684.02     &         21.5     &  3$ \times 10^{-4}$  \\

%% file: tab_speckle_obs.tex
201092629 &      562nm &  44nm &  2017-05-15 \\
201092629 &      832nm &  40nm &  2017-05-15 \\
201092629 &      562nm &  44nm &  2017-03-18 \\
201092629 &      832nm &  40nm &  2017-03-18 \\
201102594 &      562nm &  44nm &  2017-04-05 \\
201102594 &      832nm &  40nm &  2017-04-05 \\
201110617 &      832nm &  40nm &  2017-03-10 \\
201110617 &      562nm &  44nm &  2017-03-10 \\

%% file: tab_harps.tex
228801451 &  2457782.629699 &  22.960809 $\pm$  0.001844 & -0.012789 &   7.175241 & -4.5707 $\pm$   0.0098 &  0.873 &   1800.0 &      48.8 \\
201595106 &  2457782.687224 &   0.692781 $\pm$  0.002263 & -0.022588 &   6.965865 & -4.9714 $\pm$   0.0273 &  0.703 &   2400.0 &      45.0 \\
201437844 &  2457762.701586 &  -3.449696 $\pm$  0.005740 &  0.037015 &  20.649605 & -4.8647 $\pm$   0.0058 &  0.451 &   1200.0 &     101.6 \\
201437844 &  2457774.738143 &  -3.441043 $\pm$  0.005931 &  0.045533 &  20.699375 & -4.8584 $\pm$   0.0060 &  0.451 &   1800.0 &      98.7 \\
201437844 &  2457774.759707 &  -3.441611 $\pm$  0.006562 &  0.073457 &  20.632207 & -4.8629 $\pm$   0.0071 &  0.451 &   1800.0 &      90.0 \\

%% file: tab_atmospheric.tex
201092629.01 & 1.38 &     156 &           94 \\
201102594.01 & 1.38 &     128 &          318 \\
201110617.01 & 1.92 &     298 &          107 \\
201111557.01 & 1.80 &     246 &           73 \\
201127519.01 & 0.57 &     575 &         1146 \\
201128338.01 & 1.49 &      96 &           78 \\
201132684.01 & 2.08 &     194 &           37 \\
201132684.02 & 1.34 &     252 &           98 \\

%% file: tab_comparison.tex
201092629.01     &  $26.809633^{+0.001327}_{-0.001235}$     &       3.7     &  $0.0263^{+0.0011}_{-0.0007}$     &             3.0     &  $0.25^{+0.28}_{-0.17}$     &       0.4     &    $48.0^{+2.0}_{-6.0}$     &           0.7  \\
201102594.01     &   $6.513855^{+0.000534}_{-0.000660}$     &       0.0     &  $0.0656^{+0.0138}_{-0.0041}$     &             0.3     &  $0.54^{+0.37}_{-0.37}$     &       0.3     &   $23.0^{+4.3}_{-10.9}$     &           0.1  \\
201110617.01     &   $0.813175^{+0.000032}_{-0.000032}$     &       0.5     &  $0.0163^{+0.0008}_{-0.0007}$     &             0.1     &  $0.39^{+0.33}_{-0.27}$     &       0.0     &     $4.6^{+0.5}_{-1.0}$     &           0.4  \\
201111557.01     &   $2.302093^{+0.000127}_{-0.000133}$     &       0.8     &  $0.0143^{+0.0010}_{-0.0008}$     &             0.1     &  $0.40^{+0.34}_{-0.28}$     &       0.0     &    $12.0^{+1.5}_{-3.0}$     &           0.1  \\
201127519.01     &   $6.178825^{+0.000030}_{-0.000030}$     &       0.6     &  $0.1080^{+0.0024}_{-0.0016}$     &             1.1     &  $0.24^{+0.15}_{-0.16}$     &       0.3     &    $17.7^{+0.4}_{-0.8}$     &           0.6  \\
201128338.01     &  $32.652883^{+0.002143}_{-0.002309}$     &       0.6     &  $0.0418^{+0.0023}_{-0.0014}$     &             1.3     &  $0.40^{+0.32}_{-0.30}$     &       0.1     &   $57.0^{+4.8}_{-14.0}$     &           0.2  \\
201132684.01     &   $5.898463^{+0.001803}_{-0.001503}$     &       1.5     &  $0.0135^{+0.0009}_{-0.0009}$     &             0.7     &  $0.30^{+0.23}_{-0.20}$     &       0.3     &    $13.3^{+1.3}_{-2.1}$     &           0.6  \\
201132684.02     &  $10.062708^{+0.001114}_{-0.001122}$     &       1.3     &  $0.0271^{+0.0012}_{-0.0010}$     &             0.9     &  $0.43^{+0.22}_{-0.26}$     &       0.1     &    $18.9^{+1.9}_{-3.1}$     &           0.0  \\